\documentclass[12pt]{article}

\usepackage{newtxtext,newtxmath}

\usepackage{graphicx}

\usepackage[letterpaper,margin=1in]{geometry}

\renewenvironment{abstract}
	{\quotation}
	{\endquotation}

\date{}

\makeatletter
\renewcommand{\fnum@figure}{\textbf{Figure \thefigure}}
\renewcommand{\fnum@table}{\textbf{Table \thetable}}
\makeatother

\usepackage{scicite}

\usepackage{url}

\def\scititle{
The late formation of chondrites as a consequence of Jupiter-induced gaps and rings

\large Short Title: Jupiter-induced gaps and rings
}
\title{\bfseries \boldmath \scititle}

\author{
	Baibhav~Srivastava$^{1\ast\dagger}$,
	André~Izidoro$^{1\ast\dagger}$\and
	\small$^{1}$Department of Earth, Environmental and Planetary Sciences, Rice University, Houston \& 77005, USA.\and
	\small$^\ast$Corresponding author. Email: baibhav.s@rice.edu; izidoro@rice.edu\and
    \small$^\dagger$These authors contributed equally to this work.
}

\begin{document} 

\maketitle

\begin{abstract} \bfseries \boldmath
The accretion ages of the first planetesimals—the parent bodies of magmatic iron meteorites—suggest they formed within the first 0.5–1Myr of Solar System history. Yet, planetesimal formation appears to have occurred in at least two distinct phases. A temporal offset separates early-forming bodies from later-forming chondrite parent bodies, which accreted 2–3~Myr after the Solar System onset —an unresolved aspect of Solar System formation. Here we use numerical simulations to show that Jupiter’s early formation reshaped its natal protoplanetary disk. Jupiter’s rapid growth depleted the inner disk gas and generated pressure bumps and dust traps that manifested as rings. These structures caused dust to accumulate and led to a second-generation planetesimal population, with ages matching those of non-carbonaceous chondrites. Meanwhile, the evolving gas structure suppressed terrestrial embryos' inward migration, preventing them from reaching the innermost regions. Jupiter likely played a key role in shaping the inner Solar System, consistent with structures observed in Class II and transition disks.
\end{abstract}

\noindent
\subsection*{Teaser}
Jupiter sculpted rings in the young Solar System, producing late planetesimals—the source of many primitive meteorites.

\section*{Introduction}
Not all planetesimals—the building blocks of planets in the solar system—formed simultaneously. Isotopic analyses reveal an apparent spread in accretion ages, suggesting that while some planetesimals formed early, others accreted considerably later~\cite{horanetal98, bizzarroetal05, scherstenetal06}. Recent state-of-the-art planet formation models~\cite{drazkowskadullemond18,lichtenberg_bifurcation_2021,izidoro_planetesimal_2022,morbidellietal22} reproduce the early formation of planetesimals with accretion ages consistent with the parent bodies of non-carbonaceous magmatic iron meteorites, which likely formed within the first $\sim$1Myr of Solar System history\cite{Kruijeretal17,kruijer_great_2020}. In contrast, the comparatively delayed formation of non-carbonaceous chondrite parent bodies—specifically, ordinary and enstatite chondrites—which accreted $\sim$2–3Myr after the formation of calcium–aluminum-rich inclusions (CAIs; the formation of CAIs is typically considered the starting point of Solar System formation.)\cite{kitaetal05,connelyetal12,papeetal19,kruijer_great_2020}, remains poorly understood.

At face value, the late accretion of a planetesimal population does not appear readily compatible with a key feature of the solar system: its isotopic dichotomy~\cite{Kruijeretal17,kruijer_great_2020}. This dichotomy—between non-carbonaceous (NC) and carbonaceous (CC) meteorites\cite{warren11}-- is typically attributed to an early and persistent separation between inner and outer disk reservoirs, established by the formation of Jupiter~\cite{Kruijeretal17} or a pressure bump~\cite{brassermojzsis20}. In this framework, Jupiter (or a pressure bump) acts as a barrier that prevents the inward drift of pebbles from the outer disk and mixing, preserving isotopic distinctiveness. However, if this barrier was highly efficient, it would have also limited the inward delivery of small solids to the inner disk. This raises a related question: how did sufficient NC-material remain— and reaccumulate—to allow the formation of chondrite parent bodies 2–3 Myr after CAIs in the inner disk? Solids interior to Jupiter's orbit are thought to have been rapidly lost through radial drift\cite{adachietal76} or accreted into early planetesimals~\cite{izidoro_planetesimal_2022}. As a result, the formation of new planetesimals at later times is hard to reconcile with models in which the inner disk was early isolated and depleted. Sustaining enough dust for late planetesimal formation would require a long-lived or replenished reservoir, which standard disk models do not account for.

In this study, we present a state-of-the-art model that integrates several key processes in planet formation, offering a comprehensive and coherent framework for the early evolution of the Solar System. We show that the delayed accretion of chondrite parent bodies and the formation locations of the terrestrial planets are interconnected consequences of the evolving structure of the Sun’s natal protoplanetary disk, shaped by Jupiter’s early formation.

We combine planet–disk gravitational interactions, dust evolution, planetesimal formation, and planetary accretion into a unified framework to capture their coupled evolution during the early stages of planet formation. Our simulations show that Jupiter’s rapid growth depleted the inner gas disk and triggered the formation of pressure bumps and planet traps—manifesting as rings in the gas distribution. These structures created conditions favorable for the formation of a second-generation population of planetesimals from dust (``debris'') released during terrestrial planet accretion, via imperfect accretion.\cite{walsh_planetesimals_2019}

\begin{figure}
    \centering
    \includegraphics[width=0.9\linewidth]{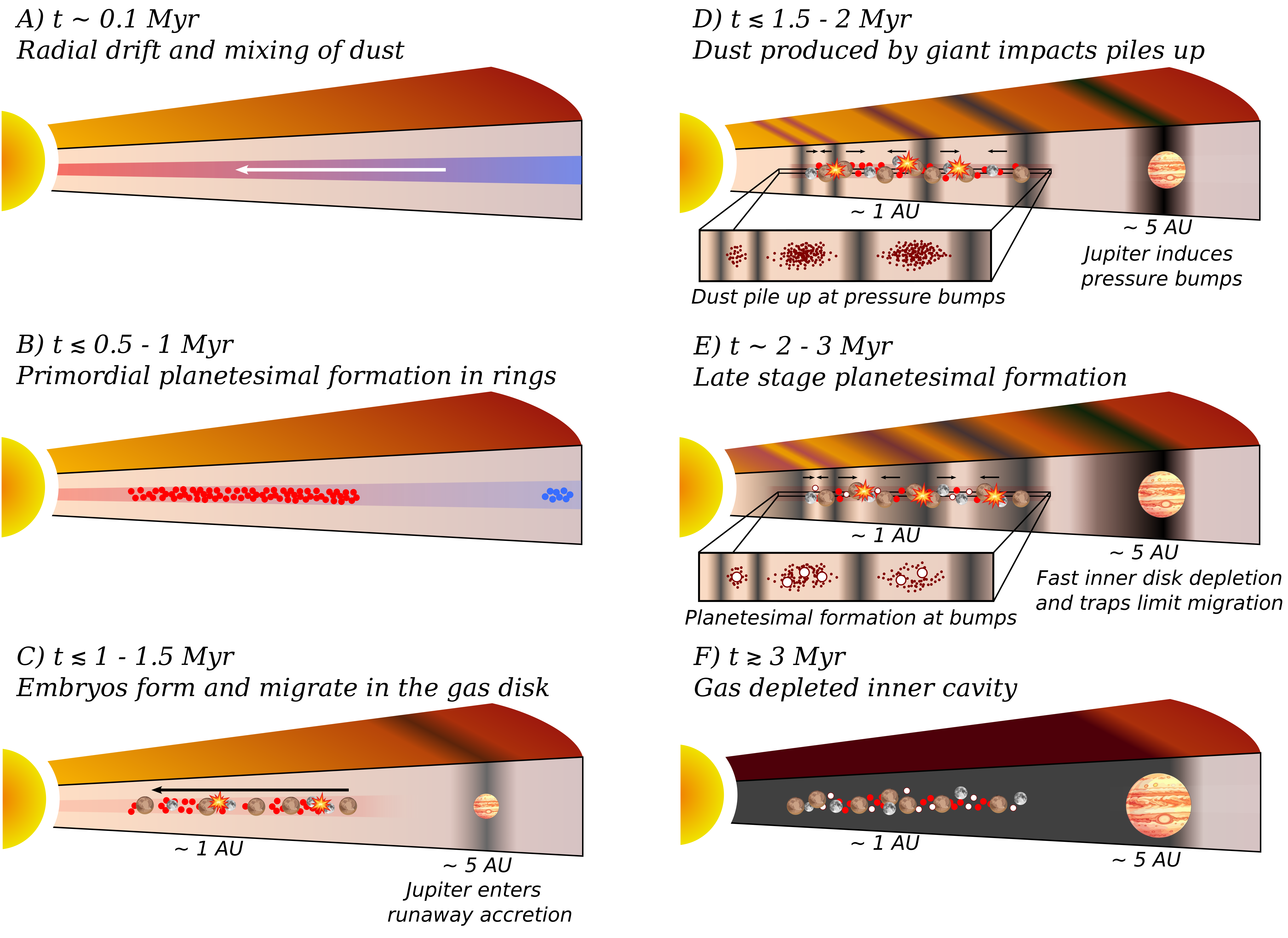}
    \caption{\textbf{Schematic illustration of the proposed evolutionary scenario for the early inner Solar System over the first $\sim$3 Myr.} (A) At early times ($t \sim 0.1$~Myr), radial drift and turbulent mixing transport dust grains across the disk.
    (B) Around $\lesssim 0.5 - 1$~Myr, primordial planetesimal formation occurs in rings\cite{izidoro_planetesimal_2022,morbidellietal22}.
    (C) By $\sim$1.5 Myr, growing planetary embryos start to migrate inward under the influence of the gaseous protoplanetary disk, while Jupiter's core enters rapid gas accretion phase.
    (D) Around $\sim$2 Myr, Jupiter’s gravitational perturbations excite spiral density waves, inducing  pressure bumps in the inner disk. Giant impacts among migrating embryos generate additional debris. Pressure bumps act as dust traps, halting inward drift of small solids and leading to dust accumulation. 
    (E) Between $\sim$2–3 Myr, dust accumulation at pressure bumps leads to the formation of a second generation of planetesimals. Rapid gas depletion in the inner disk, combined with the presence of these traps, limits the inward migration of growing embryos.
    (F) By $\sim$3 Myr, the inner gas disk is largely dissipated, resulting in a system composed of terrestrial embryos and a second generation of planetesimals—potentially the parent bodies of ordinary and enstatite chondrites—while the inner disk evolves into a gas-depleted cavity.}
    \label{fig:framework}
\end{figure}

As an outcome, our model offers a natural explanation for the orbital structure of the inner Solar System. Mars-mass objects are inferred to have formed rapidly in the terrestrial planet region — within less than 2 Myr~\cite{nimmokleine07, dauphaspoormand11} — and would have gravitationally interacted with a young and substantial gaseous disk. Given that protoplanetary disks generally persist for 3–10 Myr~\cite{mamajek09, willianscieza11, arnaudetal21}, such embryos should have undergone substantial gas-driven migration in a typical disk~\cite{linpapaloizou86, ward97a, paardekoopermellema06, ogiharaetal07, izidoro_effect_2021, woo_terrestrial_2023}. The abundance of short-period exoplanets~\cite{mayoretal11, howard12, fressin2013} appears to suggest that inward gas-driven migration may be a common outcome of planet–disk interactions~\cite{ward97a, idalin08, pierensetal11, izidoro_breaking_2017}. However, in the solar system, terrestrial planets remained concentrated around 1 au — implying that migration was more limited or even suppressed. While mechanisms such as viscosity transitions~\cite{deschturner15, broz_early_2021} and magnetically driven disk winds~\cite{ogiharaetal15, suzukietal16, woo_terrestrial_2023} have been proposed to account for their formation location, they depend on very specific and locally variable disk conditions~\cite{bai16}. In contrast, our model provides a dynamical mechanism that robustly suppresses migration via the formation of long-lived planet traps and accelerated disk dispersal linked to Jupiter’s early growth. A schematic representation of our Solar System evolution framework is presented in Figure \ref{fig:framework}

\section*{Results}
We start by modeling Jupiter's influence on its natal disk using high-resolution hydrodynamical simulations with the \textsc{\textsc{FARGO3D}} code\cite{masset_fargo_2000, benitez-llambay_fargo3d_2016}, treating the gaseous disk as a two-dimensional structure under the locally isothermal approximation. Jupiter's growth via gas accretion is not modeled in our simulations, but its mass is introduced gradually using a taper function over 800 orbits to avoid unrealistic disk perturbations\cite{hammer_slowly-growing_2017,hallampaardekooper19,lega_migration_2021}. For simplicity, Jupiter remains on a non-migrating orbit at 5.4~au throughout the simulation. This assumption is discussed later in the paper.

\begin{figure}[!ht]
	\centering
	\includegraphics[width=0.6\textwidth]{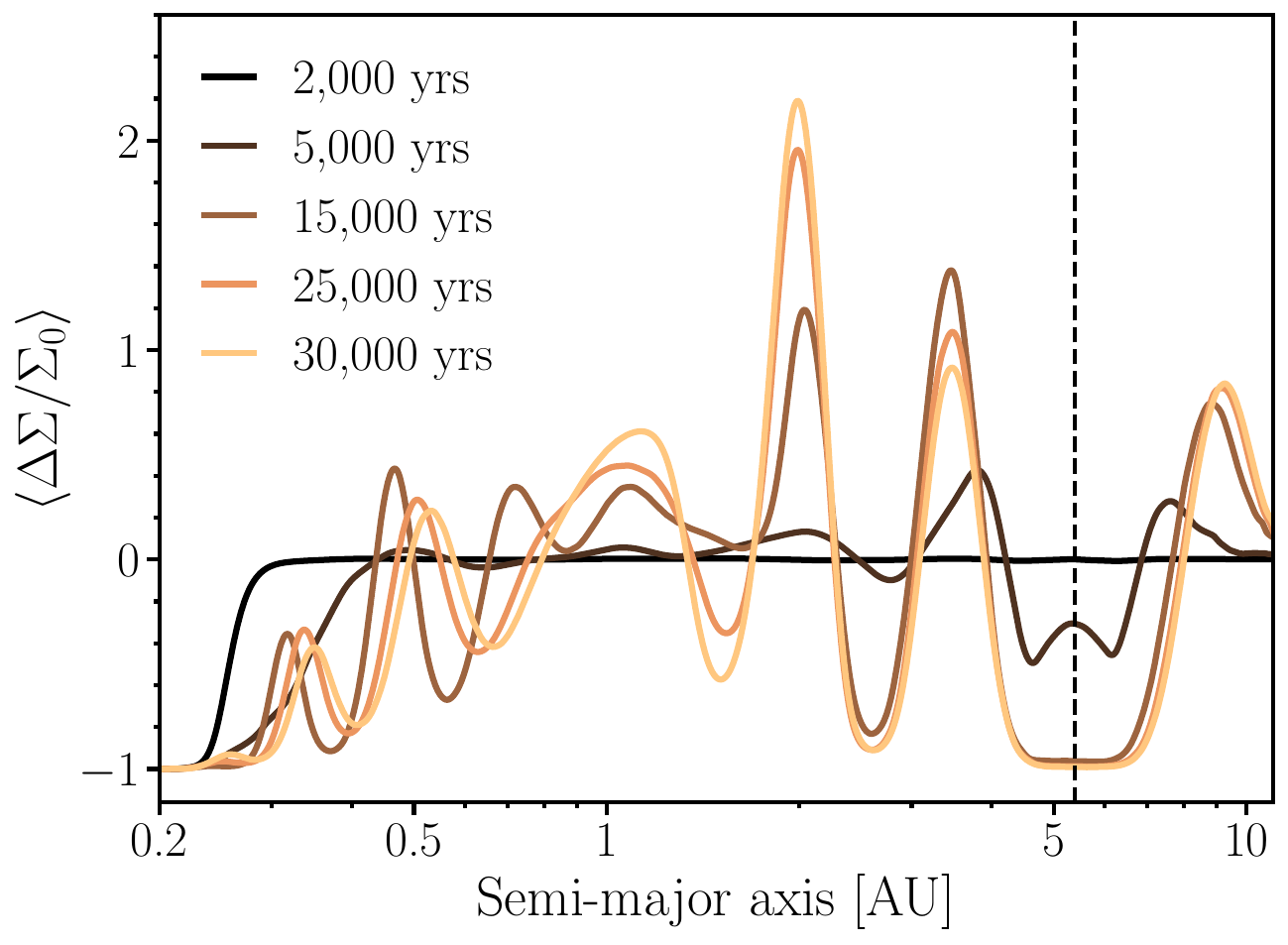}
	\caption{\textbf{Snapshots showing the formation of pressure bumps and gaps in a hydrodynamical simulation with a Jupiter-mass planet placed at 5.4~au.}
            The initial surface density profile corresponds to a power-law disk following the classical Minimum Mass Solar Nebula (MMSN) disk\cite{weidenschilling_aerodynamics_1977,hayashi81}. The disk extends from 0.2~au to 40~au, the gas viscosity is set to $\alpha=10^{-5}$, and we use free-outflow boundary conditions. During the entire course of the simulation,  Jupiter is kept in a non-migrating orbit, as indicated by the vertical dashed line. The vertical axis represents the normalized gas surface density relative to the initial surface density profile. The disk model employs a logarithmic spacing for radial resolution, consisting of 1240 radial and 1400 azimuthal cells. At 1~au, the disk resolution yields about 12 cells per scale height.}
	\label{fig:gas_surf_dens}
\end{figure}

In our study, we focus on the evolution of the inner disk, specifically the region inside Jupiter's orbit. Figure~\ref{fig:gas_surf_dens} illustrates the evolution of the normalized surface density profile in our nominal simulation, where the disk's $\alpha$-viscosity is set to $\alpha = 10^{-5}$. As expected, Jupiter opens a gap in the disk\cite{papaloizoulin84,kleynelson12} and induces the formation of multiple pressure bumps (peaks in the normalized gas density profiles) in the inner regions~\cite{bae_formation_2017,jaehanzhaohuan18}. Figure~\ref{fig:gas_surf_dens} shows that five pressure bumps form between 0.3 and 5~au. 

In relatively low-viscosity disks ($\alpha \approx 10^{-5}-10^{-4}$; see Figure~\ref{fig:gas_surf_dens} and Supplementary materials), pressure bumps form when spiral density waves excited by the planet shock the gas at specific locations\cite{goodmanrafikov01,juetal16}. These features have been observed in previous studies\cite{zhuetal13,donetal17} and have been proposed to be consistent with the ring- and gap-like structures seen in disks around young stars\cite{bae_formation_2017}. Simulations modeling this process, however,  generally employ closed-boundary conditions\cite{bae_formation_2017,jaehanzhaohuan18}, which inherently prevent the viscous evolution of the disk. These models also often focus on the outer regions, setting the inner disk edge at $r \gtrsim 1$~au\cite{bae_formation_2017}, or instead model the innermost disk as a simplified one-dimensional structure to reduce computational costs\cite{crida_cavity_2007,berguezetal23}. In our nominal simulations, the disk extends from 0.2~au to 40~au with free-outflow boundary conditions at both edges to mimic viscous evolution. The assumed inner edge at 0.2~au lies near the magnetospheric truncation radii inferred from magnetohydrodynamic simulations of star–disk interactions, which typically produce inner disk cavities at $\sim$0.1~au for T Tauri–like stars~\cite{romanova_3d_2019}. The outer edge at 40~au is aligned with some models inferring the size of the solar nebula\cite{kretke_method_2012} of about $45-50$~au. For completeness, we have verified that the pressure bumps in Figure \ref{fig:gas_surf_dens} are not artifacts of our chosen boundary conditions. 

While we present the results of our low-viscosity simulations as the nominal case, the location, number, and amplitude of pressure bumps are highly sensitive to the physical structure of the protoplanetary disk—particularly its viscosity—which we explore through a suite of simulations (see Supplementary Material). For instance, simulations with $\alpha = 10^{-3}$ do not produce strong pressure bumps inside 2~au, and those with $\alpha = 10^{-4}$ generate weaker bumps than our nominal case. Our models also adopt a locally isothermal equation of state and focus on low-viscosity regimes; however, disk thermodynamics can also play an important role in shaping pressure bump formation. In particular, disks with intermediate cooling efficiencies—where thermal relaxation times are comparable to the local dynamical time—tend to produce weaker spiral shocks, thereby reducing the amplitude of pressure maxima~\cite{rafikov02,zhuetal15,rafikov16,zhangzhu20,miranda2020planet}. By contrast, both adiabatic and fully isothermal conditions tend to support stronger and more persistent bump structures~\cite{rafikov02,miranda2020planet}.

In our nominal simulation, Figure~\ref{fig:gas_surf_dens} shows that near the disk inner edge, the gas is rapidly depleted due to accretion at the inner edge. This process is enhanced due to the transport and deposition of angular momentum into the inner regions, driven by the gravitational influence of Jupiter on the gaseous disk\cite{takeuchi_gap_1996,quillenetal04,varniere_planets_2006}. We refer to this process as ``Jupiter-induced gas accretion'' or ``shock-driven accretion'' \cite{rafikov16}). This phenomenon can be particularly important in low-viscosity disks, where the intrinsic disk viscosity is insufficient to dampen the spiral density waves excited by the planet over large radial scales\cite{takeuchi_gap_1996}. As a result,  these waves can efficiently transport angular momentum all the way from near the planet's location to the innermost regions of the disk and upon steepening into shocks, they deposit it into the disk~\cite{takeuchi_gap_1996,rafikov16,varniere_planets_2006,crida_cavity_2007}. This process is naturally more efficient for high mass planets, as gas giants\cite{rafikov02,bae_formation_2017}.

\begin{figure}[!ht]
	\centering
	\includegraphics[width=0.6\textwidth]{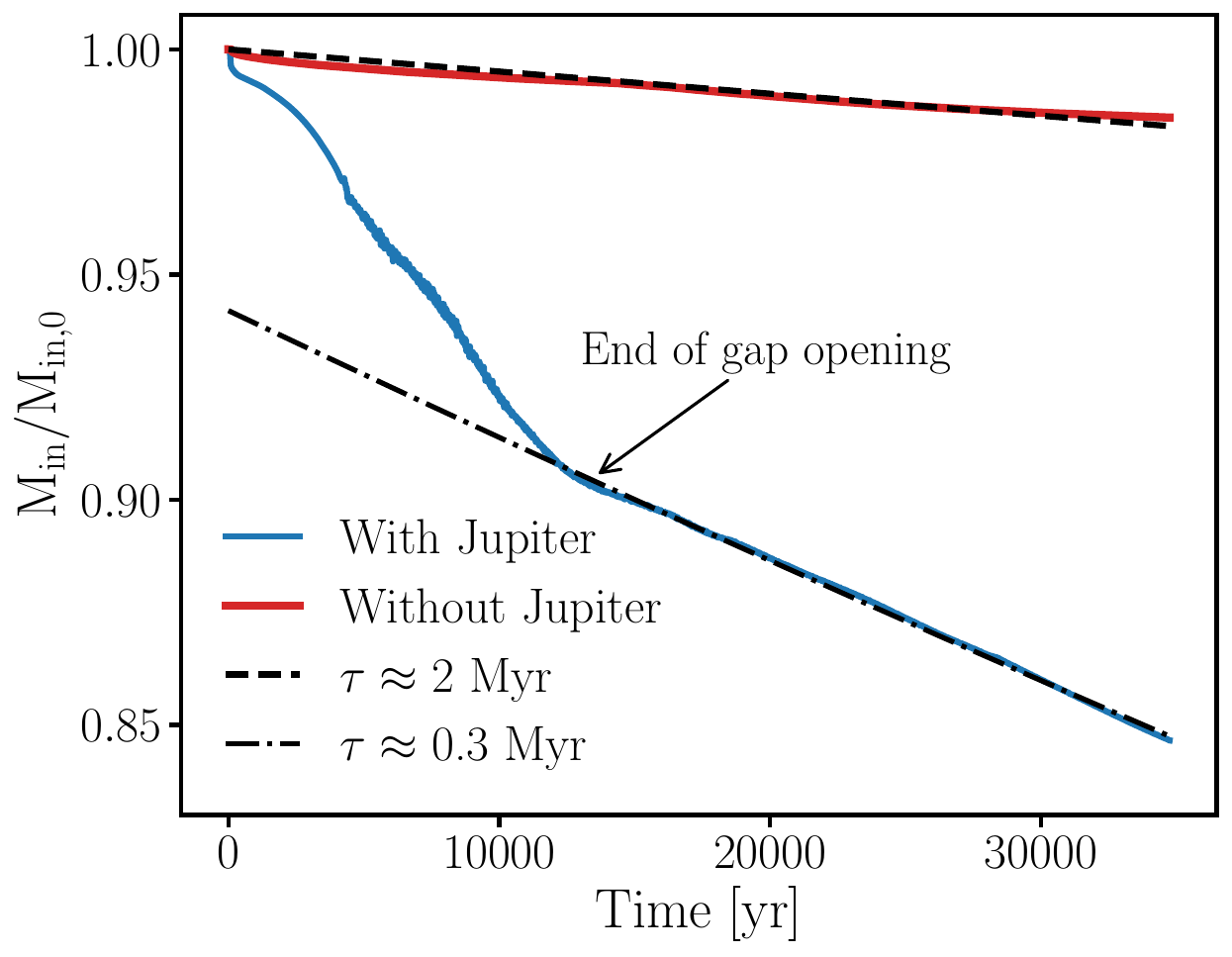}
	\caption{\textbf{Normalized inner gas disk mass over time in hydrodynamical simulations with and without Jupiter. } The vertical axis represents the mass within the inner disk ($r<5.4$~au), normalized to its initial value. The blue curve corresponds to the simulation, including Jupiter, while the red curve represents the simulation without it. The dot-dashed and dashed lines indicate the fitted exponential depletion timescales for each scenario. In the presence of Jupiter, the inner disk depletes with a timescale of 0.3~Myr --approximately seven times faster than the 2~Myr depletion timescale observed in the simulation without Jupiter.}
	\label{fig:inner_disk_mass} 
\end{figure}

Figure~\ref{fig:inner_disk_mass} compares the evolution of the inner disk mass in the simulation of Figure~\ref{fig:gas_surf_dens} and an equivalent hydrodynamical simulation without any planet. In the case including Jupiter (blue line), the inner disk mass rapidly decreases during the first 10~kyr due to gap opening by the planet and the subsequent rapid loss of gas at the disk inner edge. Once the gap-opening process is complete ($\approx10$~kyr), the mass depletion rate slightly decreases but remains consistent with a depletion timescale of approximately $\sim$0.3~Myr (the dot-dashed line represents a fitted curve proportional to $\exp(-t/\tau)$, where $\tau=0.3$~Myr). In contrast, in the absence of Jupiter (red line), the inner disk mass evolves solely due to viscous accretion, leading to a considerably longer depletion timescale of about 2~Myr (the dashed line represents a fitted curve proportional to $\exp(-t/\tau)$, where $\tau=2$~Myr), resulting in approximately a sevenfold difference between the two scenarios.

Because of the relatively low gas-disk viscosity assumed in our nominal simulation, Jupiter efficiently disconnects the ``inner'' and ``outer'' gaseous disk by opening a relatively deep gap~\cite{cridaetal06,duffelletal13,berguezetal23}. This largely prevents gas-material from the outer disk from replenishing the inner disk. We stress that some amount of gas can diffuse through Jupiter's gap (e.g.\cite{pierensetal14}) but this inflow of mass is much smaller than the flux of mass lost at the inner edge (see Supplementary Materials).  This process, along with the Jupiter's induced gas accretion (at the inner edge), lead to a relatively fast inner disk depletion. Note that, due to the high computational cost of our hydrodynamical simulations, we stopped our nominal simulation at 35~kyr. From $\sim$10~kyr to $\sim$35~kyr, the inner disk mass loss rate appears to have converged to a timescale of $\sim$0.3~Myr and remains broadly unchanged. We do not expect the depletion timescale to increase substantially beyond this point, remaining much shorter than the standard viscous timescale, until a full cavity opens in the inner disk\cite{takeuchi_gap_1996,varniere_planets_2006,crida_cavity_2007}. We now use these results to model the onset of terrestrial planet formation, incorporating the effects of Jupiter's presence on its natal disk, as demonstrated.

Our subsequent simulations model the growth of terrestrial planet embryos over 3~Myr, starting from a population of planetesimals. Planetesimals used as initial conditions in our simulations are assumed to have formed during the early stages of protoplanetary disk evolution, consistent with recent models of solar system formation \cite{izidoro_planetesimal_2022, morbidellietal22}. Although the solid-to-gas ratio in the inner disk may not have been globally favorable for planetesimal formation\cite{willianscieza11}, several disk features have been proposed to locally concentrate dust and facilitate early planetesimal formation. These include pressure bumps associated with dead zone boundaries \cite{ueda_dust_2019} and sublimation fronts \cite{drkazkowska2016close, charnozetal21}, which are thought to act as efficient dust traps. While our simulations do not model the formation of the first-generation planetesimals, we adopt initial conditions consistent with recent models of solar system formation \cite{izidoro_planetesimal_2022,morbidellietal22}, and the view that first planetesimals formed early within $\sim0.5-1$~Myr after CAIs \cite{kruijer_great_2020}.

Accordingly, we assume an initial planetesimal population distributed between $\sim$0.8 and 1.5~au. This setup is motivated by models suggesting that the terrestrial planets accreted from a narrow annulus of solids (``a ring of planetesimals'') \cite{hansen_formation_2009,drkazkowska2016close,izidoro_planetesimal_2022,morbidellietal22,woo_terrestrial_2023}, but it is naturally applicable to scenarios envisioning wide radial planetesimal distributions\cite{chambers_making_2001,walsh_planetesimals_2019}. For comparison purposes, we perform simulations including and neglecting the effects of Jupiter's formation on the inner disk. The initial gas disk scenario invoked in our planetesimal accretion simulations is the same as that used for our hydrodynamical simulations of Figure~\ref{fig:gas_surf_dens} (see Methods). We model disk dissipation by assuming an exponential decay with an e-fold timescale $\tau_{\rm pre}=2$~Myr until Jupiter starts to form. To model Jupiter’s growth within the gaseous disk over a timescale of 0.5~Myr, we linearly interpolate the disk structure between the initial power-law surface density profile and the azimuthally averaged disk profile extracted from our hydrodynamical simulations (Figure~\ref{fig:gas_surf_dens}; see Methods). In our simulations, we adopt a single timescale for Jupiter’s growth, 0.5~Myr. This parameter is not expected to substantially affect the dynamical evolution of the terrestrial planet region, and our assumed value falls within the plausible range for runaway gas accretion timescales ($\sim$0.1–1~Myr; \cite{pollacketal96})~\cite{Clementetal20}. After Jupiter's formation, the disk is dissipated using an e-fold timescale equal to $\tau_{\rm post}$, which in our nominal simulation is set to 0.3~Myr, as inferred from Figure~\ref{fig:inner_disk_mass}. We explore the effect of this parameter on our results later in the Supplementary Materials. Jupiter's formation time is treated as a free parameter in our model. In simulations neglecting the effects of Jupiter's formation, we model disk dissipation using a timescale of $\tau_{\rm }=2$~Myr during the entire simulation. 

To make our simulations modeling the onset of terrestrial planet formation computationally affordable and reasonably accurate, we use a semi-analytical model combined with N-body integration to simulate the growth of planetary embryos from 100~km planetesimals. Our model has been calibrated to broadly match the results of high-resolution N-body codes\cite{walsh_planetesimals_2019,woo_terrestrial_2023} (see Methods). We assume that our simulations start at about 0.5~Myr after the formation of CAIs, which is consistent with the accretion ages of non-carbonaceous magmatic iron meteorites. This envisions that only after about 0.5~Myr of the onset of the solar system, planetesimals formed in the terrestrial region (see \cite{lichtenberg_bifurcation_2021,izidoro_planetesimal_2022}). We account for the effects of type-I migration and tidal damping of orbital eccentricities and inclinations as planetary embryos grow. 

\begin{figure}[!ht]
	\centering
	\includegraphics[width=0.5\textwidth]{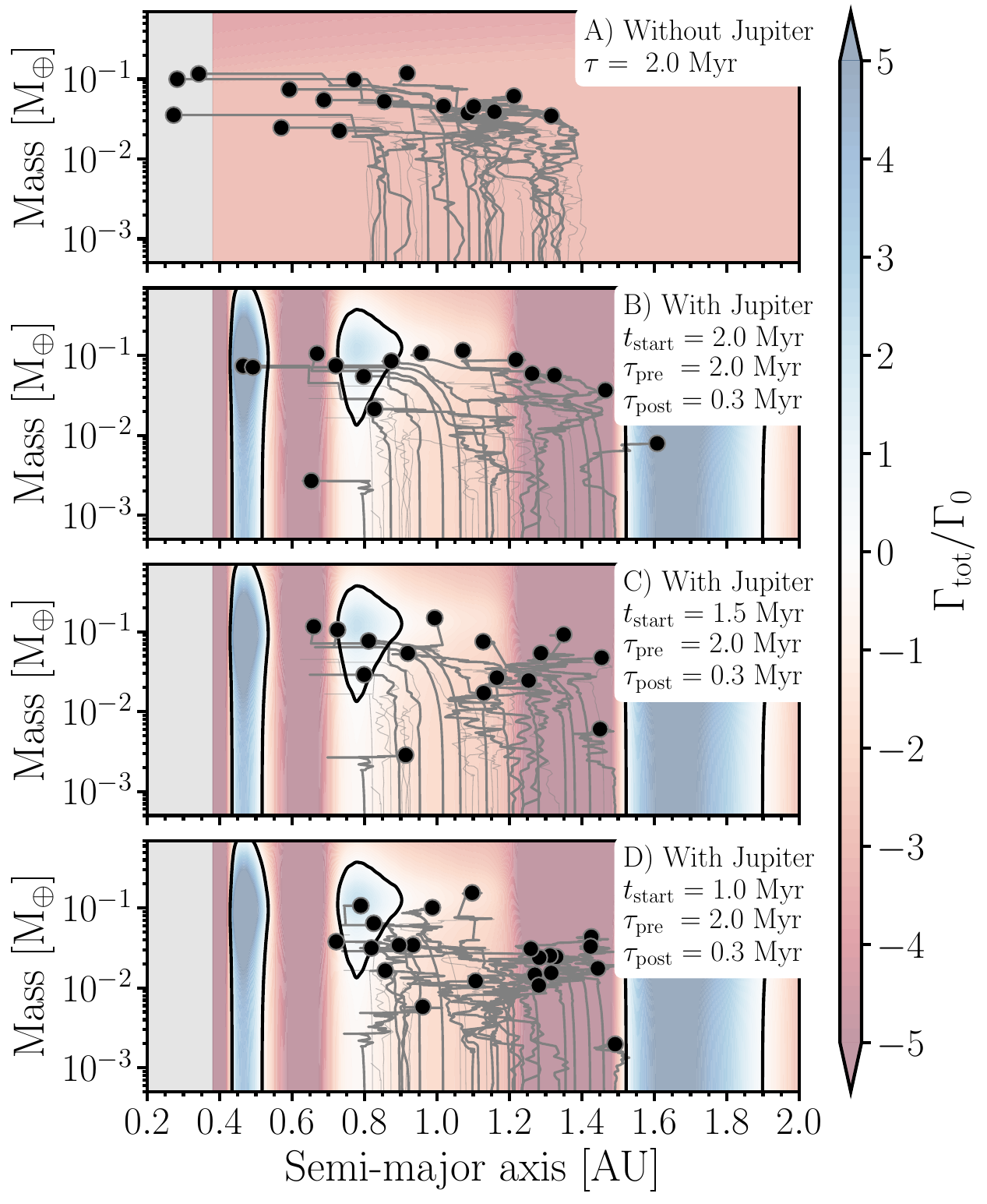} 
	\caption{\textbf{Growth-migration tracks of planetary embryos growing via planetesimal accretion and mutual impacts in simulations with Jupiter forming at varying times and without Jupiter.} Each simulation begins with 40 planetary embryos, each with a mass of $10^{-4}$ M$_\oplus$, embedded in a disk of $\sim$100~km-sized planetesimals initially distributed between 0.8 and 1.5~au (see Methods). Before Jupiter forms, gas disk dissipation follows an e-fold timescale of 2~Myr. Upon Jupiter's formation, this timescale shortens to 0.3~Myr to reflect the accelerated depletion observed in our hydrodynamical simulations. The grey lines trace the evolution of planetary embryos in mass and semi-major axis, while black circles denote their final masses and positions. The color bar indicates normalized torques acting on the embryos (a migration map): blue regions correspond to outward migration, red to inward migration, and the black contour marks zero-torque regions. The underlying migration map evolves as Jupiter grows. The contours shown here correspond to the time when Jupiter is completely formed. The four panels represent different scenarios, each corresponding to a distinct timing of Jupiter’s formation. From top to bottom: (A) a power-law disk without Jupiter, (B) Jupiter forming at 2~Myr, (C) at 1.5~Myr, and (D) at 1~Myr. The grey area marks the region inside Mercury's orbit. }
	\label{fig:mass_sma_embryos} 
\end{figure}

Figure~\ref{fig:mass_sma_embryos} shows the growth tracks of planetary embryos in four different simulations. These simulations start with about 40 planetary embryos each -- treated as gravitationally self-interacting objects-- where their individual initial masses are set to $10^{-4} M_\oplus$. In Figure~\ref{fig:mass_sma_embryos}, the growth of planetary embryos is initially dominated by accretion of 100~km sized planetesimals following the runaway/oligarchic growth regimes \cite{kokuboida96,kokuboida98}. Our initial planetesimal sizes are motivated by the outcome of streaming instability simulations\cite{simonetal16,klahr2020turbulence}. Once planetary embryos' masses grow to about moon mass, they start to gravitationally interact with the gas disk and migrate\cite{linpapaloizou86,idalin08}. As planetary embryos migrate, they may scatter each other and grow further via mutual collisions with nearby embryos. The tidal torques experienced by planetary embryos (shown by the colormap in Figure~\ref{fig:mass_sma_embryos}) depend on the local gradients of surface density and temperature profiles\cite{paardekooper_torque_2010,paardekooper_torque_2011}. Red-toned colors represent regions of inward migration (negative torque), while blue-toned colors show regions of outward migration (positive torque). It is worth noting that some embryos move outward even in the regions of inward migration, and vice versa, due to the scattering by nearby embryos. Figure~\ref{fig:mass_sma_embryos}-A shows the growth tracks of planetary embryos in a power-law disk (neglecting the effects of Jupiter's formation). In this specific case, gas-driven planet migration is mainly inward. Figures \ref{fig:mass_sma_embryos}-B, C, and D illustrate the cases in which Jupiter begins to form at 2, 1.5, and 1~Myr after CAIs, respectively. In these cases, changes in the disk structure due to Jupiter-induced pressure bumps lead to regions of outward migration. 

In general, the migration history of planetary embryos differs drastically between simulations that neglect Jupiter’s formation (power-law disks) and those that include it. In a power-law disk (Figure~\ref{fig:mass_sma_embryos}-A), as commonly used in classical models of terrestrial planet formation, planetary embryos can grow sufficiently quickly to migrate substantially inward, eventually reaching the innermost regions of the disk (inside Mercury’s orbit). This result is consistent with those of previous studies \cite{ogihara_formation_2018,izidoro_effect_2021,woo_terrestrial_2023}. Scenarios in which terrestrial planets (or their precursors) migrate inside Mercury's orbit (grey region) are inconsistent with the current architecture of the Solar System. However, in simulations where Jupiter forms sufficiently early -- typically within 2~Myr after CAIs (e.g.~Figure~\ref{fig:mass_sma_embryos}-C and D) --  the induced changes in the gaseous disk structure and the relatively fast gas depletion (after Jupiter's formation) alter the migration timescale across the inner solar system, slowing down inward planet migration in some regions and reversing its direction in others. Consequently, in Figure~\ref{fig:mass_sma_embryos}-C and D, terrestrial planetary embryos show limited inward migration, generally staying concentrated around 0.7-1~au. This is consistent with previous models modelling the late stage of accretion of terrestrial planets\cite{hansen_formation_2009,walsh_low_2011,izidoroetal15b}.

Our results depend sensitively on the assumed initial gas surface density. A higher-density disk (relative to our nominal minimum mass solar nebula model~\cite{hayashi81}) would require Jupiter to form earlier in order to produce the same dynamical effects on embryos, whereas a lower-density disk would permit a somewhat later formation. The slope of the surface density profile may also influence these outcomes\cite{paardekooper_torque_2011}. We do not explore the full parameter space of initial conditions that can influence embryo growth and gas-driven migration. In particular, the extent of migration depends not only on the disk’s structure but also on factors such as the initial solid surface density\cite{Clementetal20}, the size and mass distribution of planetesimals\cite{walsh_planetesimals_2019}, and the gas dissipation timescale~\cite{walsh_planetesimals_2019}. In our setup, we adopt initial embryo and planetesimal masses and sizes motivated by predictions from streaming instability simulations\cite{simonetal16} and semi-analytical growth models\cite{wetherill_formation_1980}, which yield bodies with masses approaching that of Mars by the time the gas disk dissipates—consistent with Mars’ inferred growth history\cite{nimmokleine07, dauphaspoormand11}. While there is an inherent trade-off between these parameters, our adopted values represent a physically reasonable choice, consistent with commonly used disk models\cite{walsh_low_2011,bitsch_structure_2015} and observational/cosmochemical constraints.

Our simulations including Jupiter's effects  show a strong planet trap at about 0.4-0.5~au, which is broadly consistent with Mercury's current location. The total mass trapped in this region depends on the timing of Jupiter’s formation. For instance, in Figure~\ref{fig:mass_sma_embryos}, when Jupiter begins to form at 2~Myr, approximately $\sim$0.1M$_\oplus$ is confined between 0.4 and 0.5AU, carried by two planetary embryos. However, embryos typically reach this region only in scenarios where Jupiter forms relatively late (see Figure\ref{fig:mass_sma_embryos}), allowing for more extensive inward migration before being halted by the trap and rapid disk dispersal. Although the presence of this trap makes it an appealing candidate for explaining Mercury’s current orbit, further work is needed to assess whether it can also account for the planet’s small mass and compositional peculiarities.

We now demonstrate that our framework is also consistent with the late accretion of the parent bodies of non-carbonaceous chondrites, which are estimated to have formed between 2 and 3 million years after CAIs\cite{kitaetal05,connelyetal12,papeetal19,kruijer_great_2020}.

During the accretion of terrestrial planets, collisions among planetary embryos and planetesimals generate dust (``collisional debris'') due to imperfect accretion\cite{kokubogenda10,chambers13}. It has been hypothesized that pressure bumps induced by a growing gas giant planet can facilitate planetesimal formation\cite{jaehanzhaohuan18}. We test this hypothesis and show that Jupiter-induced pressure bumps, coupled with the rapid depletion of the inner disk, naturally give rise to a late/second-generation planetesimal population. This population forms from dust generated by collisions among planetary embryos and planetesimals, which becomes subsequently trapped in pressure bumps in the interior of Jupiter's orbit. Furthermore, we show that the accretion ages of this planetesimal population match those of the parent bodies of non-carbonaceous chondrites.

We model dust evolution, coagulation, fragmentation, and planetesimal formation using a modified version of the \textsc{Two-Poppy} code \cite{birnstiel_simple_2012}, solving the 1D advection-diffusion equation that governs dust surface density. Our code also accounts for the effects of planetesimal formation as a sink term in the dust advection-diffusion equation (see Methods) and also dust production (an additional source term) to mimic dust created during the accretion of terrestrial planets. Dust is introduced into the simulation at a rate derived from high-resolution N-body simulations accounting for collisional evolution~\cite{izidoro_planetesimal_2022}.

\begin{figure}[!ht]
	\centering
	\includegraphics[width=\textwidth]{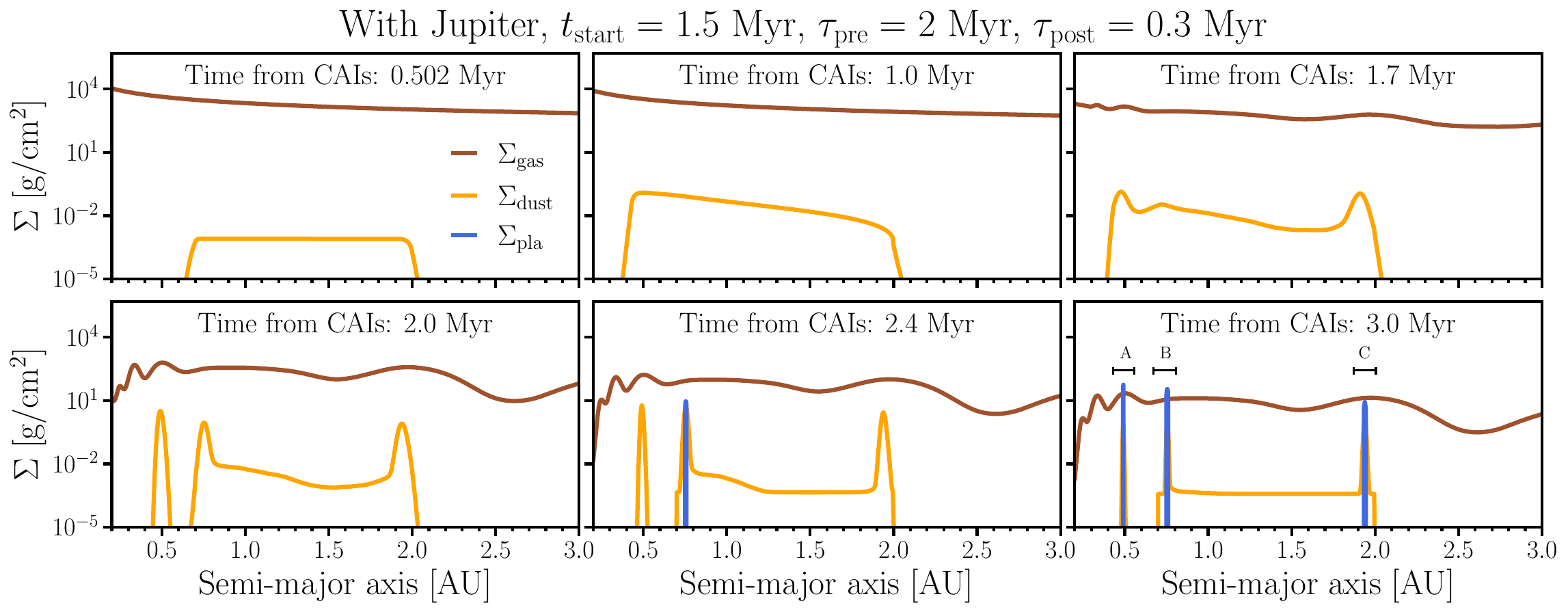} 
	\caption{\textbf{Snapshots of surface densities of gas (brown), dust (yellow), and planetesimals (blue) from a dust advection-diffusion simulation including planetesimal formation.} The initial surface density follows an MMSN-like disk profile, with no dust initially present in the disk. Dust is introduced between 0.7 and 2.0~au, as calibrated from the results of simulation modeling collisional evolution\cite{izidoro_planetesimal_2022}. The top-left panel shows the disk after 2000 years from the start of the simulation (0.502~Myr relative to CAIs) - showing a small amount of dust present. As in previous simulations, the gas disk initially dissipates following an exponential decay with a timescale of 2~Myr. Jupiter starts to form at 1.5~Myr, and we subsequently reduce the depletion timescale to 0.3~Myr. As Jupiter forms, pressure bumps appear and continue to develop until 2~Myr (see Methods). Pressure bumps facilitate dust trapping, triggering planetesimal formation starting at around 2.3~Myr. The bottom-right panel shows the final reservoirs of planetesimals -- labeled A, B, and C -- that are generated by the end of the gas disk phase.}
	\label{fig:dust_pla_surf}
\end{figure}

Figure~\ref{fig:dust_pla_surf} presents a series of snapshots illustrating the evolution of gas ($\Sigma_{\rm gas})$ and dust surface densities ($\Sigma_{\rm dust})$. As in our previous simulations, the start time of our ``dust'' evolution simulations corresponds to 0.5~Myr after the formation of CAIs. We use the same disk model used in our hydrodynamical and planetesimal accretion simulations. As dust is ``produced'', the dust surface density increases over time (compare top panels of Figure~\ref{fig:dust_pla_surf}). The gas disk initially dissipates with a depletion timescale of 2~Myr until Jupiter begins forming at $t_{\rm start}=1.5$~Myr. From that point onward, the depletion timescale shortens to $\tau_{\rm post}=0.3$~Myr (as in our nominal accretion simulations; see Figures \ref{fig:mass_sma_embryos}). Again, to mimic Jupiter’s growth within the gaseous disk, we use linear interpolation to model the emergence of Jupiter-induced pressure bumps and gaps in the gaseous disk. Figure~\ref{fig:dust_pla_surf} shows that as dust piles up at Jupiter-induced pressure bumps, planetesimals eventually start to form at about 2.3~Myr at around 0.7~au, and subsequently at 0.5 and 2~au. This leads to the formation of three narrow rings of planetesimals (``A'', ``B'', and ``C''; see Figure~\ref{fig:dust_pla_surf}). The final planetesimal surface density ($\Sigma_{\rm pla})$ is shown by the blue curves in the panel corresponding to 3~Myr. The accretion ages of this planetesimal population are consistent with estimated ages for the parent bodies of non-carbonaceous chondrites, between 2 and 3~Myr after the formation of the solar system\cite{kitaetal05,connelyetal12,papeetal19,kruijer_great_2020}.

\begin{figure}[!ht]
	\centering
	\includegraphics[width=0.6\textwidth]{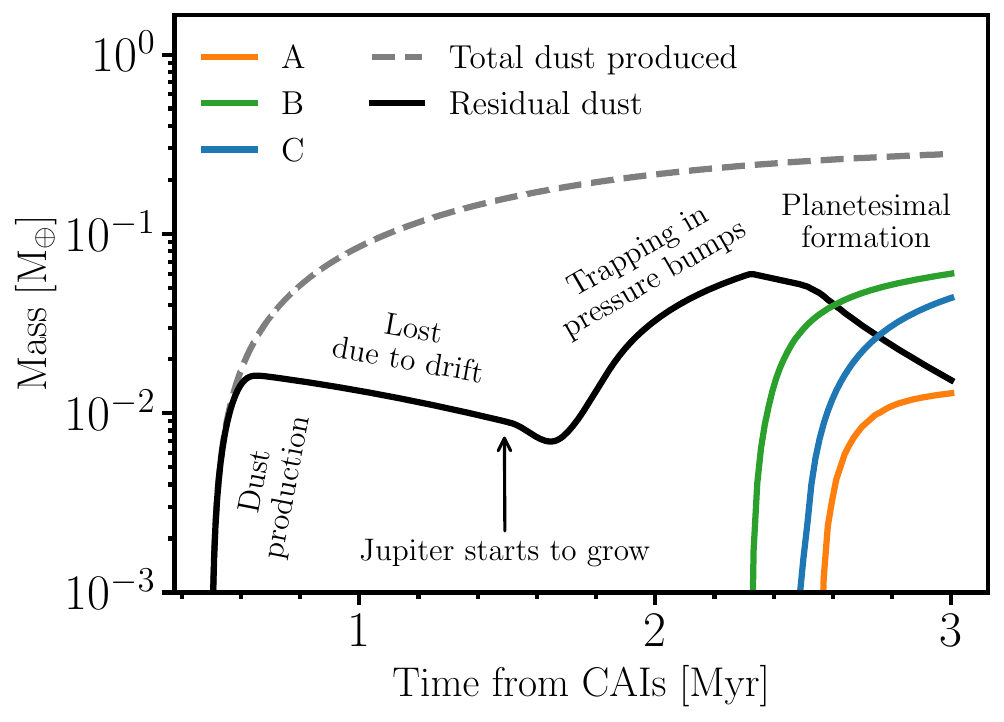} 
	\caption{\textbf{Evolution of dust and planetesimal mass over time in the simulation of Figure~\ref{fig:dust_pla_surf}.}
            The grey dashed line represents the total amount of collisional dust added to the disk, while the black line indicates the dust remaining in the disk at a given time. Jupiter begins to form at 1.5~Myr, and prior to this, most of the dust is lost to the Sun due to drift. Once Jupiter starts to grow, the dust mass increases as it accumulates in the pressure bumps. This concentration of dust leads to the formation of a second generation of planetesimals around 2.3~Myr, and this process continues until the end of the gas disk phase at 3~Myr. The orange, green, and blue lines show the masses in the three reservoirs --A, B, and C-- respectively.}
	\label{fig:chondrites_mass} 
\end{figure}

Figure~\ref{fig:chondrites_mass} shows the cumulative dust mass in the inner disk in the simulation of Figure~\ref{fig:dust_pla_surf}, along with the cumulative mass produced in planetesimals over time. The planetesimal mass in ring ``A'' is shown in orange, ``B'' in green, and ``C'' in blue. The solid black line shows the existing dust mass as a function of time in the inner disk. As dust is added, it is also simultaneously lost due to radial drift (between 1 and $\sim$1.5~Myr). When Jupiter begins to grow, which in Figure~\ref{fig:chondrites_mass} occurs at $t_{\rm start}=1.5$~Myr, pressure bumps form and begin to trap dust (see also Figure~\ref{fig:dust_pla_surf}). The concentration of dust in pressure bumps ultimately leads to conditions where gravitational instabilities take place\cite{youdin_planetesimal_2002,johansenyoudin07}, leading to planetesimal formation\cite{simonetal16,carreraetal22} (see Methods). 

\begin{figure}[!ht]
	\centering
	\includegraphics[width=0.7\textwidth]{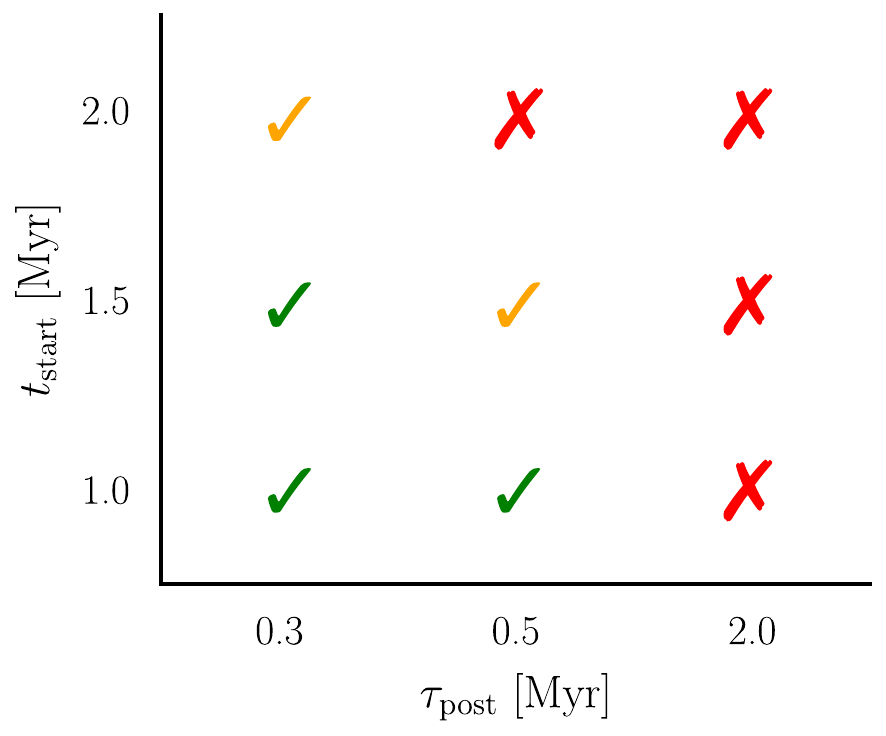}
	\caption{\textbf{Summary of the outcome of simulations modeling dust coagulation and planetesimal formation with different Jupiter formation times ($\mathbf{t_{\rm start}}$; relative to CAIs) and post-formation disk dissipation timescales ($\mathbf{\tau_{\rm post}}$)}. The green checkmarks represent parameter sets where planetesimals successfully formed in the inner disk. We assume that a ``successful scenario'' forms  $\gtrsim 10^{-1}$ M$_\oplus$ in planetesimals between 2 and 3~Myr. The yellow checkmarks are ``maybe'' scenarios where $\sim 10^{-2}$ M$_\oplus$  in planetesimals were produced between 2-3~Myr. The red crosses are the parameter sets where planetesimals were not formed in the first 3~Myr.}
	\label{fig:summary}
\end{figure}

The total mass of planetesimals produced in our simulations depends on Jupiter’s formation time and the rate at which the disk dissipates after its formation. In Figure~\ref{fig:chondrites_mass}, where Jupiter forms at 1.5~Myr and the disk dissipates on a timescale of $\tau_{\rm post} = 0.3$~Myr, the final total mass of planetesimals at 3~Myr is about 0.2~$M_{\oplus}$. In simulations where Jupiter forms relatively later, at 2~Myr after CAIs, while the gaseous disk dissipates at the same rate ($\tau_{\rm post} = 0.3$~Myr), only 0.02~$M_{\oplus}$ planetesimals form. This reduction occurs because a larger fraction of dust is lost before Jupiter’s formation (see Figure~\ref{fig:summary}). Conversely, in simulations where Jupiter begins forming at 1.5~Myr, but the gaseous disk depletes slowly over $\tau_{\rm post} = 2$~Myr, no planetesimals form in the inner disk during 3~Myr, as the necessary conditions for planetesimal formation are not met (the gas disk is not sufficiently depleted; see Supplementary Materials). 

In our nominal simulations, planetesimals are typically formed in three or four distinct rings, as in Figure~\ref{fig:dust_pla_surf} (see also Supplementary Materials). We speculate that their different accretion regions may provide an avenue for explaining the different isotopic compositions\cite{kruijer_great_2020,burkhardtetal21} and oxidation states\cite{schaeferfegley07,grewal_accretion_2024} seen in different types of non-carbonaceous chondrites, such as Enstatite and Ordinary chondrites. Notably, our model does not rule out the possibility that additional chondrite-forming reservoirs may have existed~\cite{burkhardtetal21}. Some of these could have been fully accreted into planets or are underrepresented in our current meteorite collections due to sampling biases or collisional destruction. Thus, while the formation of three/four distinct planetesimal populations is appealing, it likely does not preclude the existence of other regions of second-generation planetesimal formation in the early solar system.

Importantly, most of the planetesimals in our simulations form in rings located outside the main asteroid belt, requiring subsequent implantation in the belt\cite{raymond_origin_2017,bottkeetal06}. Dynamical models show that the implantation efficiency of planetesimals inside 2 au into the asteroid belt ranges from $\sim10^{-4}$-$10^{-2}$\cite{raymondizidoro17,izidoro_planetesimal_2022}. The total mass in S-complex type asteroids in the main asteroid belt is $1.05\times10^{-4}M_{\rm \oplus}$ \cite{demeocarry14}. For an implantation efficiency of $\sim10^{-2}$ at least $\sim0.01M_{\rm \oplus}$ in planetesimals is required to account for the entire S-complex population (if the belt was born completely empty\cite{raymondizidoro17}). In our scenarios yielding the most favorable results, the total mass of planetesimals is typically $\gtrsim0.1$$M_{\oplus}$, which would require an implantation efficiency of $\sim10^{-3}$. We summarize our results exploring different Jupiter's formation times and  $\tau_{\rm post}$ in Figure~\ref{fig:summary}. Additional simulations exploring different dust distributions and Jupiter’s formation times are presented in the Supplementary Materials.

\section*{Discussion}
Our simulations demonstrate that Jupiter’s induced rapid inner gaseous disk depletion, gaps, and rings are broadly consistent with both the birthplaces of the terrestrial planets and the accretion ages of the parent bodies of non-carbonaceous chondrites. 
Our results suggest that Jupiter formed early, within $\sim 1.5$–$2$~Myr of the Solar System's onset, and strongly influenced the inner disk evolution. Our model favors a low-viscosity regime for the Sun's natal disk ($\alpha \approx 10^{-5} - 10^{-4}$ or lower), broadly consistent with the range of viscosities inferred for observed young disks \cite{isellaetal16,dullemondetal18}.

Like any numerical model, our simulations involve simplifications. For instance, we do not explicitly model the formation of chondrules --millimeter-sized droplets formed from flash-heated molten materials. Ordinary and enstatite chondrites contain a substantial fraction of chondrules, along with a complementary fine-grained matrix \cite{alexanderetal08}. The origins of chondrules remain debated\cite{deschetal12,cieslaetal13}, with impact-generated formation processes \cite{asphaugmovshovitz11,sandersscott12,wakita17} considered viable hypotheses. 

In our simulations, we also neglect the effects of Jupiter's gas-driven migration \cite{ward97a}. This simplification is motivated by the fact that, once Jupiter opens a deep gap in a low-viscosity disk, its migration is expected to be fairly slow, particularly as the inner disk becomes depleted. Simulations show that, in low-viscosity disks, migration can be halted or reversed depending on the local disk structure \cite{monschetal21,lega_migration_2021}. In reality, Jupiter probably formed beyond the initial position assumed in our simulations and first migrated via type-I migration and eventually entered in the type-II regime (for our disk with $\alpha=10^{-5}$ a deep gap is open when the mass reaches about 10$M_{\oplus}$; \cite{kanagawaetal15}), but its exact migration history is difficult to constrain~\cite{mcnally2019migrating,lega_migration_2021}. Additionally, we do not account for Saturn’s formation and influence on the disk.  It is well established that Jupiter and Saturn typically undergo convergent migration, forming a resonant configuration that can halt or reverse their migration direction \cite{massetsnellgrove01,pierensetal14}. If Jupiter (and Saturn) migrate, the locations of pressure bumps in the disk may evolve over time. However, our assumption that Jupiter remains near its final orbit during the gas disk phase offers a reasonable first-order approximation, particularly given the slow and potentially self-limiting nature of giant planet migration in low-viscosity disks.

Our findings offer a coherent explanation for the apparent tension between the solar system’s isotopic dichotomy and the late formation of ordinary and enstatite chondrites. By demonstrating that Jupiter’s early growth can deplete the inner gas disk while simultaneously generating pressure bumps and dust traps, our model allows for the retention and local regeneration of solid material in the inner disk—enabling the formation of a second generation of planetesimals. This framework can naturally preserve the NC–CC isotopic separation while accounting for the delayed accretion ages of non-carbonaceous chondrite parent bodies, thereby linking cosmochemical constraints with the evolving dynamical structure of the disk.

Our model suggests that Jupiter must have begun forming within approximately 1.5–2 Myr of the onset of Solar System formation. CC-like asteroids in the asteroid belt are thought to have originated in the outer solar system and to have been implanted during the growth and migration of the giant planets \cite{raymond_origin_2017}. A potential caveat to this framework is that, as Jupiter grew, it likely scattered nearby planetesimals inward \cite{raymond_origin_2017}. If these planetesimals were predominantly carbonaceous (CC), as traditionally assumed, this could imply early delivery of CC material to the inner solar system—posing an apparent contradiction with the observed isotopic dichotomy between NC and CC reservoirs. This caveat is not unique to our model but any framework where Jupiter starts to form early\cite{kruijer_great_2020}. We propose two possible resolutions to this issue. First, Jupiter may have undergone runaway gas accretion in a locally depleted region—such as a gap between two planetesimal rings \cite{izidoro_planetesimal_2022}—or after clearing most planetesimals from its feeding zone. Alternatively, Jupiter may have formed in a distinct, non-carbonaceous (NC) reservoir beyond the snowline, separate from the NC reservoir that sourced the terrestrial planets. This latter scenario is particularly compelling in light of recent evidence that some of the oldest NC planetesimals (e.g., magmatic iron meteorites) accreted with substantial water content, consistent with formation beyond or near the snowline \cite{grewal_accretion_2024}. In this context, Jupiter would have primarily scattered NC-rich planetesimals inward, mixing them with more reduced inner disk material. Regardless of the specific scenario, the delivery of CC material into the asteroid belt appears to be a late-stage process ($\gtrsim$3 Myr) —likely driven by the growth and migration of Saturn or the ice giants \cite{izidoroetal15,desouzaetal22}, during the final stages of inner disk dispersal when the local gas surface density had fallen to just a few percent of its initial value \cite{raymond_origin_2017}, while the outer disk likely remained gas-rich, as suggested by observations of transition disks \cite{vandermarel23}.

\subsection*{Materials and Methods}

\subsubsection*{Hydrodynamical simulations}

We performed two-dimensional hydrodynamical simulations using \textsc{FARGO3D} \cite{masset_fargo_2000, benitez-llambay_fargo3d_2016}, adopting a locally isothermal equation of state and including a Jupiter-mass planet embedded in the disk. Our simulations were performed considering different disk $\alpha$-viscosities\cite{shakura_black_1973} (Supplementary Materials). In our nominal simulations, we use $\alpha=10^{-5}$. 

We used free-outflow open boundary conditions at the inner and outer edges of the disk to mimic viscous evolution and disk depletion. The radial grid in our nominal simulation spans from 0.2 to 40~au. The chosen inner edge location is motivated by disk cavity locations observed in magnetohydrodynamical simulations \cite{romanova_3d_2019}, while the outer edge location aligns with the disk size suggested by the inclinations and eccentricities of Kuiper Belt Objects \cite{kretke_method_2012}. We adopted logarithmic radial spacing for our simulations.

To assess convergence in our hydrodynamical simulations, we conducted a series of tests with the inner edge set at 0.5~au using closed boundary conditions. We evaluated three different resolutions: 36 cells per scale height at 1~au ($N_r$ = 2654, $N_a$ = 5190, where $N_r$ and $N_a$ represent the number of radial and azimuthal cells, respectively), 24 cells per scale height at 1~au ($N_r$ = 1724, $N_a$ = 3370), and 12 cells per scale height at 1~au ($N_r$ = 824, $N_a$ = 1640). The resulting surface density profiles exhibited convergence within 5\% across these resolutions. Given the substantial computational cost of high-resolution simulations, we selected a resolution of 12 cells per scale height at 1~au for our nominal simulations with open boundary conditions. Running a single simulation, such as the one shown in Figure~\ref{fig:gas_surf_dens}, requires approximately 6.5 months of dedicated computing time on 24 cores. Our simulations with closed boundary conditions (Supplementary materials) were conducted at a resolution of 24 cells per scale height at 1~au.

The initial gas surface density in our simulations follows a simple power-law profile \cite{weidenschilling_aerodynamics_1977,hayashi81}:
\begin{equation}
    \Sigma_\text{MMSN}(r) = \Sigma_0\left(\frac{r}{\text{1~au}}\right)^{-1}, \label{eq:powlaw}
\end{equation}

\noindent\ where $\Sigma_0= 1700{\rm g/cm^2}$.

The disk aspect ratio is 0.05 at 1~au, with a flaring index of 0.25:
\begin{equation}
    h(r) = 0.05\left(\frac{r}{\text{1~au}}\right)^{0.25}
\end{equation}

In our simulations, we neglect Jupiter’s radial migration and gas accretion, and instead fix its orbit at 5.4au. This choice is motivated by the expectation that a Jupiter-mass planet embedded in a low-viscosity disk undergoes slow radial evolution, primarily through the Type II migration regime\cite{lega_migration_2021,monschetal21}. Rapid introduction of the planet’s gravitational potential can artificially induce perturbations and unphysical vortices~\cite{hammer_slowly-growing_2017,hallampaardekooper19}. To mitigate these effects, we gradually initiate Jupiter’s potential over 800 orbital periods using the standard mass-tapering function implemented in \textsc{FARGO3D}.

\subsubsection*{Gas disk profile}
We simulated planetesimal growth, gas-driven planet migration, and dust dynamics using the gas surface density profile from our nominal hydrodynamical simulation. In order to mimic Jupiter's growth in the gaseous disk, we use linear interpolation between the initial power-law gas disk profile and the final output of our hydrodynamical simulation (Figure~\ref{fig:gas_surf_dens}).
The evolution of the gas  surface density is divided into three regimes:
\begin{equation}
    \Sigma_\text{gas}(r, t) = \begin{cases}
        \Sigma_\text{MMSN}(r)\exp\left(-\frac{t}{\tau_\text{pre}}\right), & t < t_\text{start} \\
        \Sigma_\text{bumpy}(r, t)\exp\left(-\frac{t}{\tau_\text{post}}\right), & t_\text{start} \leq t < t_\text{end} \\
        \Sigma_\text{final}(r)\exp\left(-\frac{t}{\tau_\text{post}}\right), & t \geq t_\text{end}, \label{eq:regimes}
    \end{cases}
\end{equation}

\noindent\ where $t_\text{start}$ is the time when Jupiter starts to grow\cite{pollacketal96}, $t_\text{end}$ is the time when Jupiter reaches its final mass, $\tau_\text{pre}$ is the depletion timescale for $t < t_\text{start}$, $\tau_\text{post}$ is the depletion timescale for $t \geq t_\text{start}$, $\Sigma_\text{final}$ is the final surface density profile from our hydrodynamical simulation (Figure~\ref{fig:gas_surf_dens}). In Eq. \ref{eq:regimes}, $\Sigma_\text{bumpy}(t)$ is the linearly interpolated gas density profile, given by:
\begin{equation}
    \Sigma_\text{bumps}(r, t) = \Sigma_\text{MMSN}(r) -  \frac{\Sigma_\text{MMSN}(r) - \Sigma_\text{final}(r)}{t_\text{end} - t_{\text{start}}}\left(t - t_\text{start}\right).
\end{equation}
$\Sigma_\text{MMSN}$ is the power-law profile given by equation \ref{eq:powlaw}. In all simulations, Jupiter grows within a timescale of 0.5~Myr\cite{pollacketal96}, and hence $t_\text{end} = t_\text{start} + 0.5$~Myr. $t_\text{start}$ is a free parameter in our model,  ranging from 0.5 to 2~Myr. Our nominal depletion timescales are consistent with the results of Figure~\ref{fig:inner_disk_mass}. It should be noted that $t = 0$ in the above equations corresponds to 0.5~Myrs from CAIs. 

\subsubsection*{Growth of planetary embryos from planetesimals}

We use a semi-analytical approach to compute the growth of planetary embryos via planetesimal accretion \cite{izidoro_effect_2021}. Our simulations start with 40  objects of $10^{-4} M_\oplus$ each, referred to as ``planetary embryos", embedded in a disk of $\sim$ 100 km-sized planetesimals. Embryos are allowed to grow by accreting planetesimals and also via mutual collisions. In the limit when the gravitational stirring and gas-drag damping are in equilibrium, the root-mean-squared eccentricities and inclinations of the planetesimals are given by \cite{kokubo_formation_2000, thommes_oligarchic_2003, ida_scattering_1993}:
\begin{equation}
    e_\text{pla} = 2i_\text{pla} = 2.7\left(\frac{R_\text{pla}\rho_\text{bulk}}{\Delta C_da_\text{Emb}\rho_\text{g}}\right)^{1/5}\frac{r_\text{mH}}{2^{1/3}a_\text{Emb}}
\end{equation}

\noindent\ where $R_\text{pla}$ is the radius of the planetesimals, $\rho_\text{bulk}$ is the bulk density of the planetesimals, $a_\text{Emb}$ is the semi-major axis of the growing embryo, $\rho_\text{g}$ is the volume density of the gas, $r_\text{mH} = a_\text{Emb}\left(\frac{2M_\text{Emb}}{3M_\odot}\right)^{1/3}$ is the mutual Hill radius of adjacent growing embryo, $C_d$ is the drag-coefficient and $\Delta$ is the separation of adjacent embryos in Hill radii \cite{kokubo_formation_2000}. The bulk density of the planetesimals is set to 3 g/cm$^3$. 

The accretion rate of embryos is given by\cite{wetherill_formation_1980}:
\begin{equation}
    \frac{dM_\text{Emb}}{dt} = F\frac{\Sigma_\text{pla}^{\star}\pi R_\text{Emb}^2}{\sin(i_\text{pla})a_\text{Emb}\sqrt{2}}\times\left(1 + \frac{v_\text{esc}^2}{v_\text{rel}^2}\right)v_\text{rel}
    \label{eq:dmdt}
\end{equation}
where $\Sigma^{\star}_\text{pla}$ is the local planetesimal surface density, $R_\text{Emb}$ is the radius of the embryo, $v_\text{esc} = \sqrt{\frac{2GM_\text{Emb}}{R_\text{Emb}}}$ is the escape velocity of the embryo and $v_\text{rel}$ is the relative velocity between the planetesimals and the growing embryo. $v_\text{rel}$ is given by \cite{spaute_accretional_1991}:
\begin{equation}
    v_\text{rel} = v_\text{K}\sqrt{\frac{5}{8}e_\text{pla}^2 + \frac{1}{2}(\sin i_\text{pla})^2}.
\end{equation}

\noindent\ In Equation \ref{eq:dmdt}, $F$ is a fudge factor introduced to account for using the root-mean-squared values. We set $F = 3$ following previous studies~\cite{greenzweig_accretion_1992,izidoro_effect_2021}.

Embryos and planetesimals are initialized between 0.8 and 1.5~au\cite{izidoro_planetesimal_2022,woo_terrestrial_2023}. The initial total mass in planetesimals is $\sim 2M_\oplus$. This scenario is consistent with recent solar system formation models that propose an early separation between the inner and outer solar system reservoirs \cite{izidoro_planetesimal_2022,morbidellietal22}. In particular, a pressure bump near the snowline—possibly associated with sublimation fronts—may have acted as an early barrier to the inward drift of solids, even before the formation of Jupiter \cite{brassermojzsis20,charnozetal21,izidoro_planetesimal_2022}. Motivated by this framework, our planetesimal accretion simulations begin at 0.5 Myr after CAIs with a narrow annulus of solids in the inner disk containing approximately two Earth masses, and we neglect the inward drift of pebbles from the outer disk. If solids from the outer disk were also considered to drift inward and contribute to planetesimal formation or pebble accretion in this inner ring, the total mass in solids would increase—likely resulting in terrestrial planets that are too massive if the ring exceeds about two Earth masses\cite{izidoro_planetesimal_2022,morbidellietal22}. While one could, in principle, reduce the initial mass to compensate, doing so would require fine-tuning the pebble flux and planetesimal formation efficiency to ultimately reproduce the observed mass budget\cite{izidoro_planetesimal_2022}. Instead, we build on previous studies and focus on the dynamical and accretional evolution of the inner disk once the non-carbonaceous (NC) reservoir has formed and become largely decoupled from the outer solar system.

In the inner ring, as planetary embryos grow by accreting planetesimals, $\Sigma^{\star}_\text{pla}$ is updated at every timestep to account for respective local depletion. We neglect the radial diffusion of planetesimals due to embryo-planetesimal scattering and migration of embryos. This semi-analytical approach allows us to model the growth of planetary embryos starting from planetesimal-sized objects without having a large number of self-interacting bodies in the N-body simulation. We have verified that our approach leads to results broadly consistent with other N-body codes\cite{walsh_planetesimals_2019,woo_terrestrial_2023}. The total mass in leftover planetesimals at the end of the gas disk phase in our model is about $\sim1~M_{\oplus}$, typically 50\% of the initial total planetesimal mass. The remaining mass is carried by planetary embryos. 

\subsubsection*{Gas-driven planetary migration}
We also account for the effects of the tidal interaction of growing planetary embryos with the gaseous disk. Planet-disk gravitational interactions lead to gas-driven planet migration\cite{ward97a,paardekoopermellema06}. The total torque experienced by a growing embryo is modeled in our work as
\begin{equation}
    \Gamma_\text{tot} = \Delta_\text{L}\Gamma_\text{L} + \Delta_\text{C}\Gamma_\text{C},
    \label{eq:torque}
\end{equation}

\noindent\ where $\Gamma_\text{L}$ and $\Gamma_\text{C}$ are the Lindblad and corotational torques. In Eq. \ref{eq:torque}, $\Delta_\text{L}$ and $\Delta_\text{C}$ are torque reductions due to embryo's orbital eccentricity and inclination \cite{cresswell_three-dimensional_2008,coleman_formation_2014,fendyke_corotation_2014}. The torque components depend on the local gradients of density and temperature of the disk. We compute the torques using the formulas of Paardekooper et al.\cite{paardekooper_torque_2010,paardekooper_torque_2011} for the locally isothermal regime. We also account for the effects of saturation of the Lindblad and Corotation torques due to viscous diffusion\cite{paardekooper_torque_2011}, and orbital eccentricity and inclination damping\cite{cresswell_three-dimensional_2008}. Simulations were performed using a modified version of the N-body code \textsc{Mercury}\cite{chambers_hybrid_1999,izidoro_effect_2021}. We model collisions as perfect merging events that conserve mass and linear momentum. We also recall that dust (``collisional dust'') is introduced in our simulations using a parametrization derived from high-resolution simulations\cite{izidoro_planetesimal_2022}.

\subsubsection*{Dust dynamics and planetesimal formation}

We use a modified version of the publicly available code Twopoppy \cite{birnstiel_simple_2012} to model dust evolution and planetesimal formation. The gas disk surface density is the same as that used in our accretion simulations. 
An additional source term is added to the 1D advection-diffusion equation to model the production of dust, while planetesimal formation appears as a sink term:
\begin{equation}
    \frac{\partial\Sigma_\text{dust}}{\partial t} + \frac{1}{r}\frac{\partial}{\partial r}\left\{r\left[\bar{v}_{r,\text{dust}}\Sigma_\text{dust} - D\frac{\partial}{\partial r}\left(\frac{\Sigma_\text{dust}}{\Sigma_\text{gas}}\right)\Sigma_\text{gas}\right]\right\} + \frac{\partial\Sigma_\text{dust,coll}}{\partial t} = \frac{\partial\Sigma_\text{pla}}{\partial t},
\end{equation}
 
\noindent\ where $\frac{\partial\Sigma_\text{dust, coll}}{\partial t}$ is the collisional dust production rate obtained from fits of N-body simulations \cite{izidoro_planetesimal_2022}, $\bar{v}_{r,\text{dust}}$ is the mass-weight radial velocity of the dust component, and $D$ is the dust diffusivity. When solving for the dust velocity, we account for the gas advection.

The mass evolution of collisional dust is given by the following parametric equation:
\begin{equation}
    \frac{M_d(t)}{M_{\oplus}} = a\tanh{\left(b\;\frac{t}{\rm Myr}\right)}
\end{equation}
where $a=0.32$ and $b=0.55$ are the best-fit parameters.

At the time zero of our simulations, we assume $\Sigma_\text{dust} = \Sigma_\text{pla} = 0$ everywhere in the simulation domain. Collisional dust is added to the simulation only between 0.7 and 2.0~au, a region slightly broader than the region where terrestrial planetary embryos are growing, to account for some level of diffusion and stirring. Previous simulations estimated that the total amount of dust produced during the accretion of terrestrial planets varies from about 0.2 to 0.4~${\rm M}_{\oplus}$\cite{izidoro_planetesimal_2022}.

In our simulations, planetesimal formation is triggered once the volume density of the largest dust grains  ($\rho_\text{dust}$) exceeds the volume density of the gas ($\rho_\text{gas}$) at the mid-plane\cite{drkazkowska2016close,drkazkowska2017planetesimal}:
\begin{equation}
    \frac{\partial\Sigma_\text{pla}}{\partial t} = \begin{cases}
        -\varepsilon\frac{\Sigma_\text{dust}}{T_k}, & \text{if } \frac{\rho_\text{dust}}{\rho_\text{gas}} \geq 1 \text{ and } \text{St} \geq 10^{-3} \\
        0,&\text{otherwise},
    \end{cases}
\end{equation}

\noindent\ where $T_k = 2\pi/\Omega_k$ is the orbital period and $\varepsilon$ is the planetesimal formation efficiency, which is set to $10^{-5}$\cite{izidoro_planetesimal_2022} for our nominal simulations. We also assume that pebbles are only converted to planetesimals if their Stokes number is greater than a critical value of $10^{-3}$\cite{drazkowskadullemond18,ueda_dust_2019}. The Stokes number is given by:
\begin{equation}
    \text{St} = \frac{\pi a \rho_\text{bulk}}{2\Sigma_\text{gas}}
\end{equation}

\noindent\ where $a$ is the size of dust and $\rho_\text{bulk}$ is the bulk density of dust, which is set to 3 g/cm$^3$.

In our nominal simulations modeling dust evolution and planetesimal formation, we assume that dust particles initially have millimeter sizes and that their size does not evolve with time, motivated by the typical chondrule sizes observed in non-carbonaceous chondrite meteorites\cite{nelsonrubin02,friedrichetal15}. For completeness, we also explore a scenario in which dust grains start as micrometer-sized particles and grow to their maximum size, which is determined by local disk conditions, including midplane turbulence, gas density, and the fragmentation velocity of silicate grains, as in standard dust coagulation models\cite{birnstiel_simple_2012,drazkowskadullemond18}. In this second scenario,  dust particles are also allowed to undergo fragmentation as they dynamically evolve in the gaseous disk. This approach tracks both the largest and the smallest dust grains, assuming that most of the dust mass is carried by the largest particles\cite{birnstiel_simple_2012}. Both scenarios lead to broadly similar results.

In our nominal dust coagulation simulations, we have neglected the sink term associated with dust accretion by planetary embryos. However, we have verified that a typical Mars-mass embryo located at a pressure maximum can accrete up to $\sim$50\% of the local dust mass from 2 to 3 Myr, provided its orbital inclination remains sufficiently low~\cite{johansenlambrechts17}. This estimate is based on an idealized scenario in which the embryo is dynamically cold and remains fixed at the center of a narrow dust trap—typically only $\sim$0.05~au wide. In more realistic conditions, higher eccentricities, inclinations, lower-mass, or radial offsets from the trap center (due to scattering) would considerably reduce accretion efficiency. Even if the global reduction is about 90\%, the results of our nominal simulations remains qualitatively valid.

\newpage


\clearpage 

\bibliography{jupiter}
\bibliographystyle{sciencemag}


\section*{Acknowledgments}
We are very grateful to the anonymous reviewers for their thoughtful feedback. We are also thankful to  Sean Raymond, Sho Shibata, Mengrui Pan, Sanskruti Admane, Camille Bergez-Casalou, Bertram Bitsch,  and Seth Jacobson for their stimulating and helpful discussions. 

\paragraph*{Funding:}
This work was supported in part by the Big-Data Private-Cloud Research Cyberinfrastructure MRI award funded by NSF under grant CNS-1338099 and by Rice University's Center for Research Computing (CRC).

\paragraph*{Author contributions:}
B.~S. performed the numerical simulations, curated data, carried out formal analysis, implemented software, prepared the figures and visualizations in discussion with A.~I., and contributed to validation. A.~I. conceived the project, developed the methodology, and conducted preliminary hydrodynamical simulations. He also secured resources and funding, supervised and administered the project, curated data, contributed software, and participated in formal analysis, visualization, and validation. The manuscript was written jointly by A.~I. and B.~S., with both authors contributing to the original draft and to review and editing.

\paragraph*{Competing interests:}
There are no competing interests to declare.

\paragraph*{Data and materials availability:}
Hydrodynamical simulations presented here were performed using a publicly available version of the code \textsc{FARGO3D} publicly available at \url{https://fargo3d.bitbucket.io/intro.html}. N-body simulations were performed using modified versions of the \textsc{Mercury} N-body integrator\cite{chambers_hybrid_1999}, publicly available at \url{https://github.com/4xxi/mercury}. Dust coagulation and planetesimal formation simulations were performed using a modified version of the code \textsc{Two-poppy} that is also publicly available at \url{https://github.com/birnstiel/two-pop-py}. All simulation data supporting the findings of this study are available at \url{https://doi.org/10.5281/zenodo.16071586}.


\subsection*{Supplementary materials}
Supplementary Text\\
Figs. S1 to S6
\newpage

\newpage


\renewcommand{\thefigure}{S\arabic{figure}}
\renewcommand{\thetable}{S\arabic{table}}
\renewcommand{\theequation}{S\arabic{equation}}
\renewcommand{\thepage}{S\arabic{page}}
\setcounter{figure}{0}
\setcounter{table}{0}
\setcounter{equation}{0}
\setcounter{page}{1} 

\clearpage
\begin{center}
\section*{Supplementary Materials for\\ \scititle}

Baibhav~Srivastava$^{1\ast\dagger}$,
André~Izidoro$^{1\ast\dagger}$\\
\small$^\ast$Corresponding author. Email: baibhav.s@rice.edu; izidoro@rice.edu\\
\small$^\dagger$These authors contributed equally to this work.
\end{center}

\subsubsection*{This PDF file includes:}
Supplementary Text\\
Figures S1 to S6\\

\newpage


\subsection*{Supplementary Text}

\subsubsection*{The effects of viscosity}

\begin{figure}[!ht]
	\centering
	\includegraphics[width=0.43\textwidth]{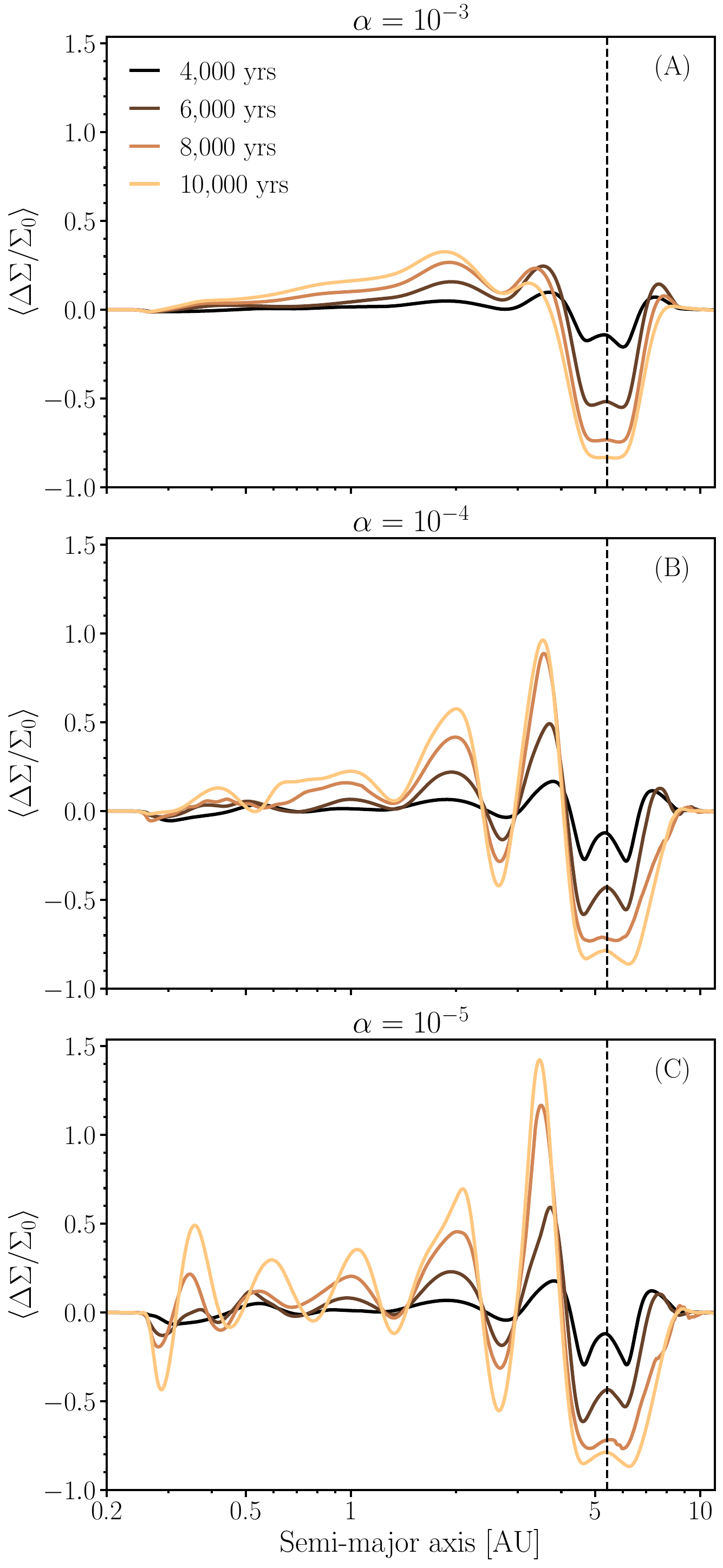}
	\caption{\textbf{Snapshots of normalized gas surface density in hydrodynamical simulations with a Jupiter-mass planet placed at 5.4 au in disks with different $\mathbf{\alpha}$-viscosity parameters.}
            Similar to our nominal simulations (e.g., Figure \ref{fig:gas_surf_dens}), these simulations also start from a power-law disk profile following the MMSN model, but we instead use closed boundary conditions with damping. Jupiter is kept in a non-migrating orbit, as indicated by the dashed line. The three panels illustrate the evolution of normalized surface density for disks with different viscosities: $\alpha = 10^{-3}$ (A), $\alpha = 10^{-4}$ (B), and $\alpha = 10^{-5}$ (C). These simulations were performed with a resolution of 24 cells per scale height at 1~au.}
	\label{fig:varying_visc}
\end{figure}

The disk viscosity plays an important role in the formation and location of pressure bumps (rings in the gas distribution) in the disk \cite{bae_formation_2017}. A higher disk viscosity results in greater damping of the spiral waves generated by the planet as they propagate, which in turn limits the number of pressure bumps that can form. Conversely, a low-viscosity disk allows the spiral arms to propagate farther from the planet, leading to shocks in the disk and the formation of multiple pressure bumps\cite{bae_formation_2017,jaehanzhaohuan18}. In figure \ref{fig:varying_visc}, we see the evolution of surface densities for three different disk $\alpha$ viscosities: $10^{-3}$, $10^{-4}$, and $10^{-5}$. These simulations are similar to the nominal one but utilize closed boundary conditions instead. As the viscosity decreases, pressure bumps are observed to form farther from the planet. Note that the observed differences in bump amplitudes in  Figure \ref{fig:gas_surf_dens} and figure \ref{fig:varying_visc} arise from differences in the simulation time and gas surface density (due to different boundary conditions). More importantly, figure \ref{fig:varying_visc} confirms that the five bumps seen in Figure \ref{fig:gas_surf_dens} of the main paper are not artifacts caused by specific boundary conditions.   As stated in the main paper, our scenario favors a low-viscosity regime for the Sun's natal disk, specifically  $\alpha\lesssim10^{-4}$.

\begin{figure}[!ht]
	\centering
	\includegraphics[width=\textwidth]{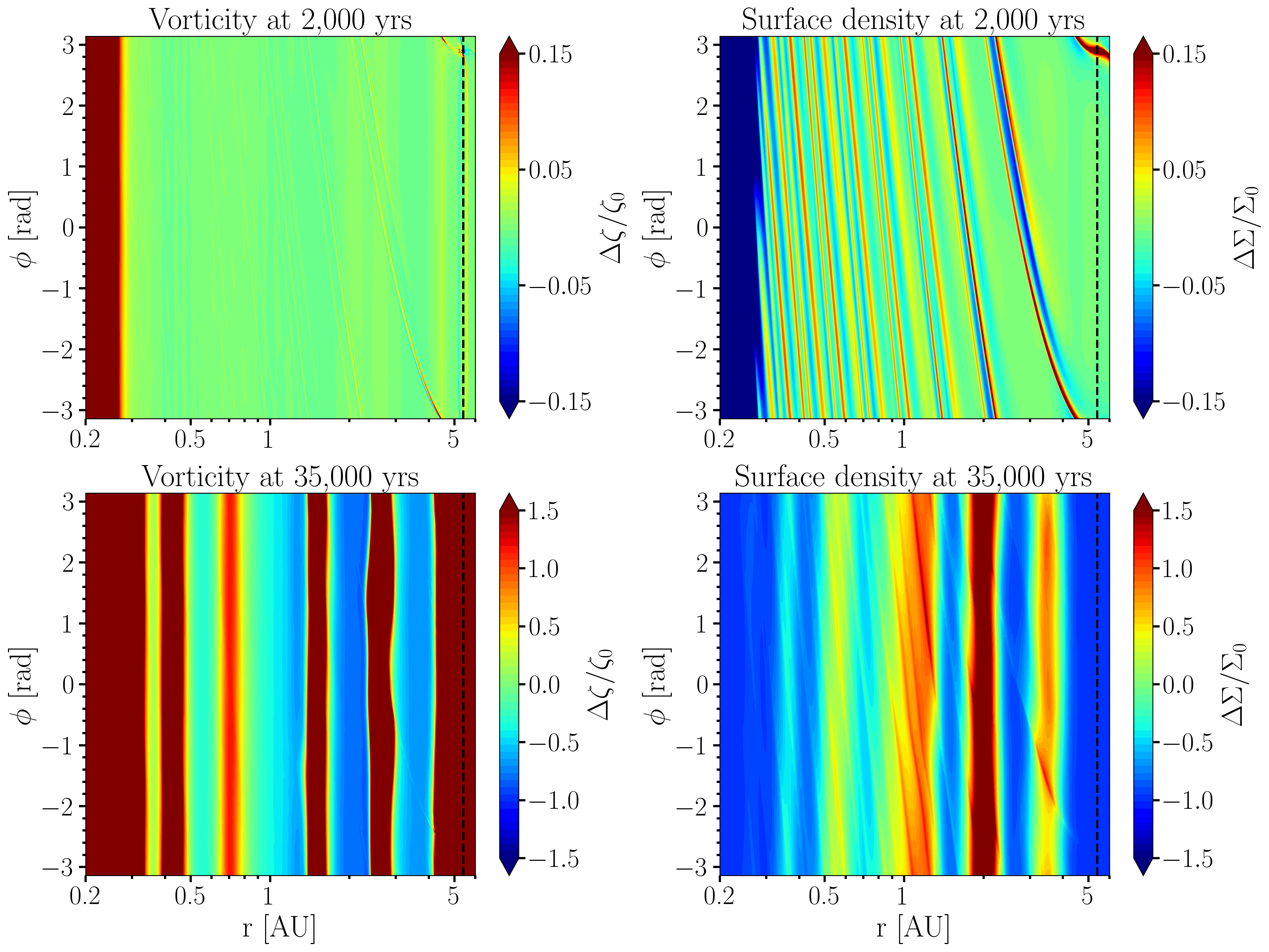}
	\caption{\textbf{Normalized potential vorticity and surface density of the disk at 2,000 and 35,000 years of our nominal simulation.} This simulation is the same as that shown in Figure \ref{fig:gas_surf_dens} of the main paper. The left panels display the disk’s potential vorticity, while the right panels show the surface density. The top and bottom rows correspond to different evolutionary times: 2,000 years (top) and 35,000 years (bottom). The x-axis represents the radial direction, and the y-axis corresponds to the azimuthal direction. Both vorticity and surface density are normalized to their initial values to enhance the visibility of fine structures.}
	\label{fig:vort}
\end{figure}

Supplementary figure \ref{fig:vort} shows the potential vorticity and the normalized surface density at 200 and 3700 orbits in the simulation of Figure \ref{fig:gas_surf_dens} (main paper). As seen, the locations where pressure bumps form correspond to specific locations where the gas shocks, as indicated by the disk potential vorticity. A high normalized potential vorticity indicates the location of a gap.

\newpage
\subsubsection*{Dust production throughout the inner disk}

\begin{figure}[!ht]
	\centering
	\includegraphics[width=\textwidth]{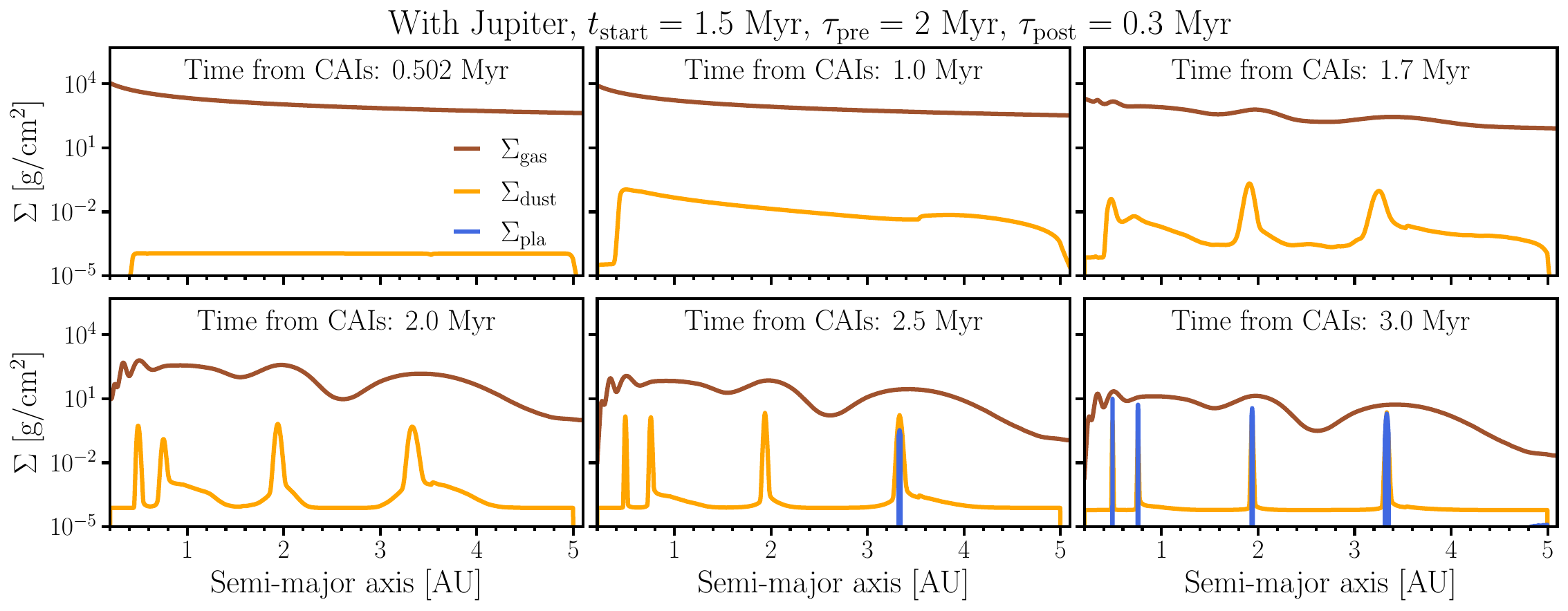}
	\caption{\textbf{Snapshots of surface densities of gas, dust, and planetesimals from a dust advection-diffusion simulation incorporating planetesimal formation, with dust production throughout the inner disk} The setup for this simulation is similar to that shown in Figure~\ref{fig:dust_pla_surf}, with the key difference being that dust is initially introduced into a wider region, extending from 0.2 to 5~au. Jupiter begins to form at $t_{\rm start}=1.5$~Myr, creating pressure bumps and accelerating the inner gas disk depletion timescale to 0.3~Myr. Pressure bumps trap dust and ultimately lead to the formation of planetesimals in four distinct reservoirs.}
	\label{fig:two_rings_dust_everywhere}
\end{figure}

In our nominal simulations, we assume that during the growth of terrestrial planets through planetesimal accretion, dust is produced exclusively between 0.7 and 2~au -- a region largely overlapping with the zone where terrestrial planetary embryos are forming (see our nominal simulations and \cite{hansen_formation_2009,izidoro_effect_2021}). In figure~\ref{fig:two_rings_dust_everywhere}, however, we adopt a more generous distribution, initializing and adding dust throughout the entire inner disk. In Supplementary figure~\ref{fig:two_rings_dust_everywhere}, dust is trapped in four out of five pressure bumps in the inner disk, ultimately leading to planetesimal formation in all of them.

\subsubsection*{Growth-tracks for planetary embryos with 2~Myr gas disk depletion timescale}

\begin{figure}[!ht]
    \centering
    \includegraphics[width=0.6\linewidth]{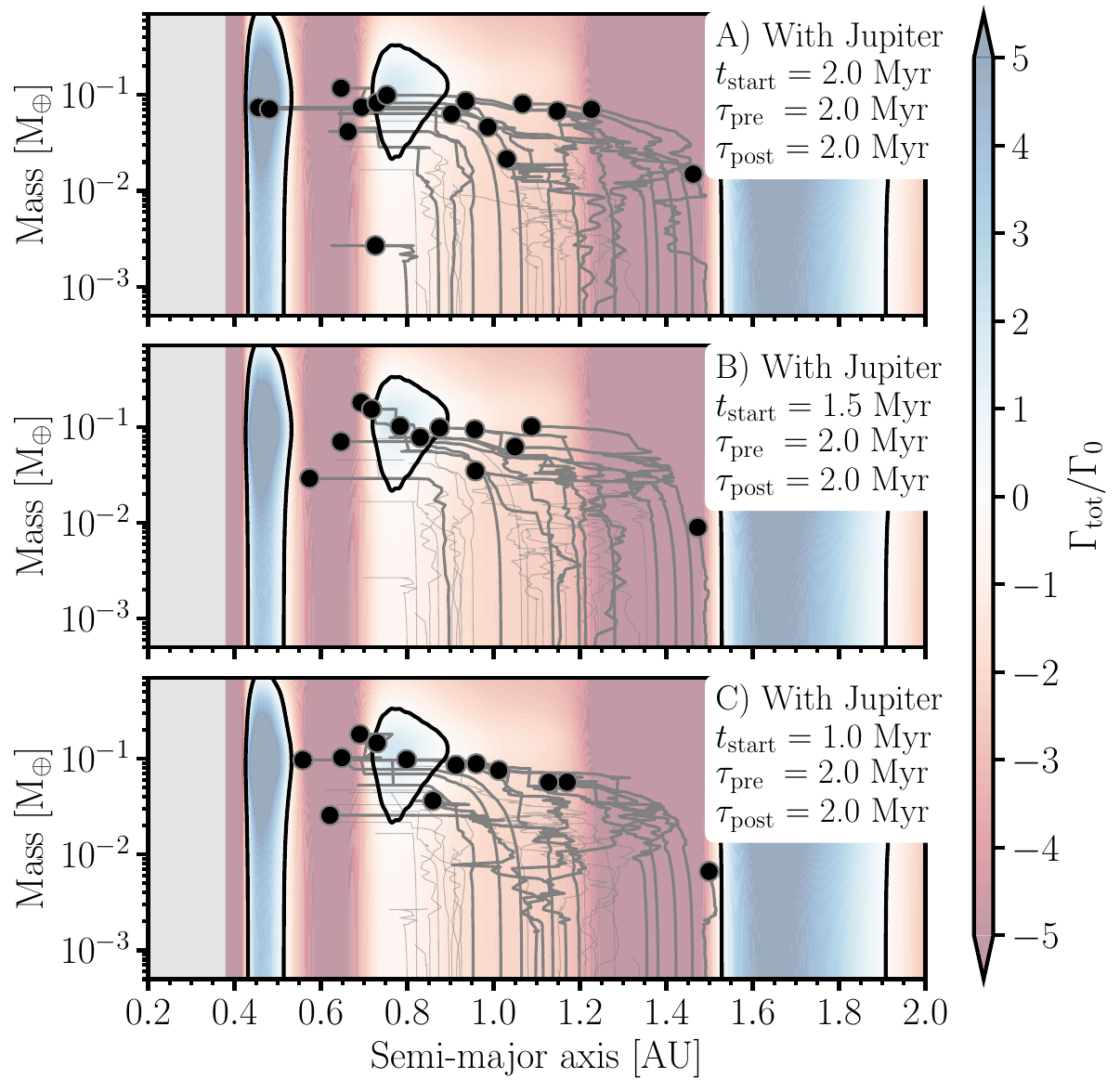}
    \caption{\textbf{Growth-migration tracks of planetary embryos in simulations with Jupiter forming at varying times and with 2~Myr depletion timescale for gas density} The setup of the simulation is similar to the nominal simulations presented in Figure~\ref{fig:mass_sma_embryos}. However, here, the gas disk depletion timescale remains fixed at 2~Myr during the entire course of the simulation and does not shorten after Jupiter's formation ($\tau_{\rm pre}=\tau_{\rm post}=2$~Myr). The three panels show three different scenarios for the timing of Jupiter's formation. From top to bottom, panels A), B), and C) show simulations in which Jupiter begins to grow at 2, 1.5, and 1~Myr, respectively.}  
    \label{fig:mass_sma_embryos_no_dep}
\end{figure}

To evaluate the impact of the gas disk depletion timescale on the growth and migration of planetary embryos, we performed N-body simulations similar to those discussed in the main text but with a fixed gas disk depletion timescale of 2~Myr throughout the simulation. The results of these simulations are presented in Supplementary figure~\ref{fig:mass_sma_embryos_no_dep}.

Our results show that even without the relatively rapid gas depletion observed in our nominal simulations (e.g.~$\tau_{\rm post}=0.3$~Myr) following Jupiter's formation, the evolution of terrestrial planet embryos remains qualitatively unchanged when Jupiter forms within 1.5~Myr. The migration of growing planetary embryos is slowed down by planet traps created by Jupiter, leading to the accumulation of planetary embryos, primarily around 0.8~au.

\subsubsection*{Effects of gas depletion on the formation of late planetesimals}

\begin{figure}[!ht]
	\centering
	\includegraphics[width=\textwidth]{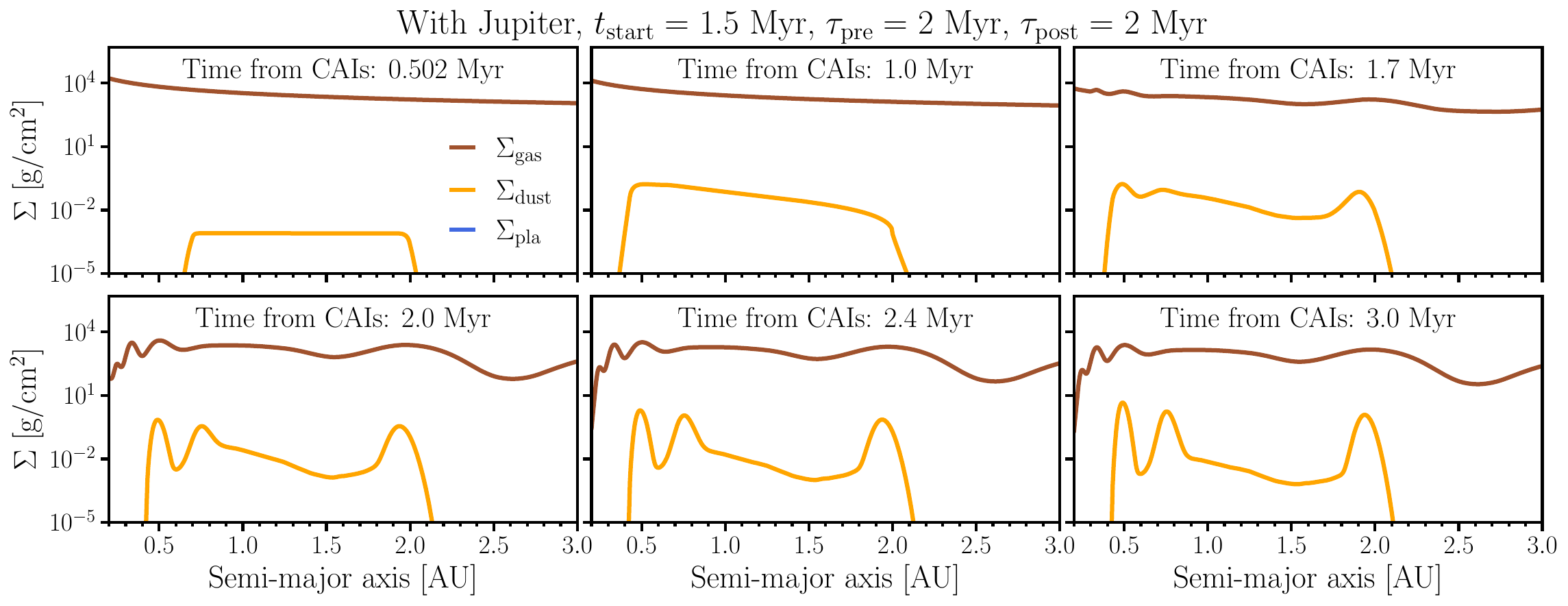}
	\caption{\textbf{Snapshots of surface densities of gas, dust, and planetesimals from a dust advection-diffusion simulation incorporating planetesimal formation with 2~Myr depletion timescale for gas density} The setup of this simulation is the same as shown in Figure~\ref{fig:dust_pla_surf}, with one key difference being that the disk dissipation timescale does not change after the formation of Jupiter, specifically $\tau_{\rm pre} = \tau_{\rm post} = 2$~Myr. As Jupiter begins to form at 1.5~Myr, pressure bumps lead to the accumulation of dust. However, planetesimal formation does not occur in the disk, as the gas densities remain too high and planetesimal formation conditions are not met.}
	\label{fig:no_dep_dust}
\end{figure}

In the main paper, we investigated the role of accelerated gas depletion in the formation of late-stage planetesimals. To further illustrate its importance, we present the results of one of our dust diffusion-advection simulations that neglects the accelerated gas disk depletion induced by Jupiter. In this scenario, Jupiter forms at 1.5~Myr and induces pressure bumps in the disk, but the gas disk depletes at a constant timescale of 2.0~Myr throughout the simulation. The result of this simulation is shown in figure~\ref{fig:no_dep_dust}. 

As in the nominal simulation, dust begins to accumulate in the pressure bumps after 1.5~Myr. However, due to the slower gas depletion, the local dust-to-gas ratio never exceeds unity, preventing the formation of planetesimals within the first 3~Myr. These findings indicate that both the presence of pressure bumps and the accelerated depletion of the gas disk are crucial for the formation of the second generation of planetesimals.

\subsubsection*{Inner and outer gas disk}

\begin{figure}[!ht]
    \centering
    \includegraphics[width=\linewidth]{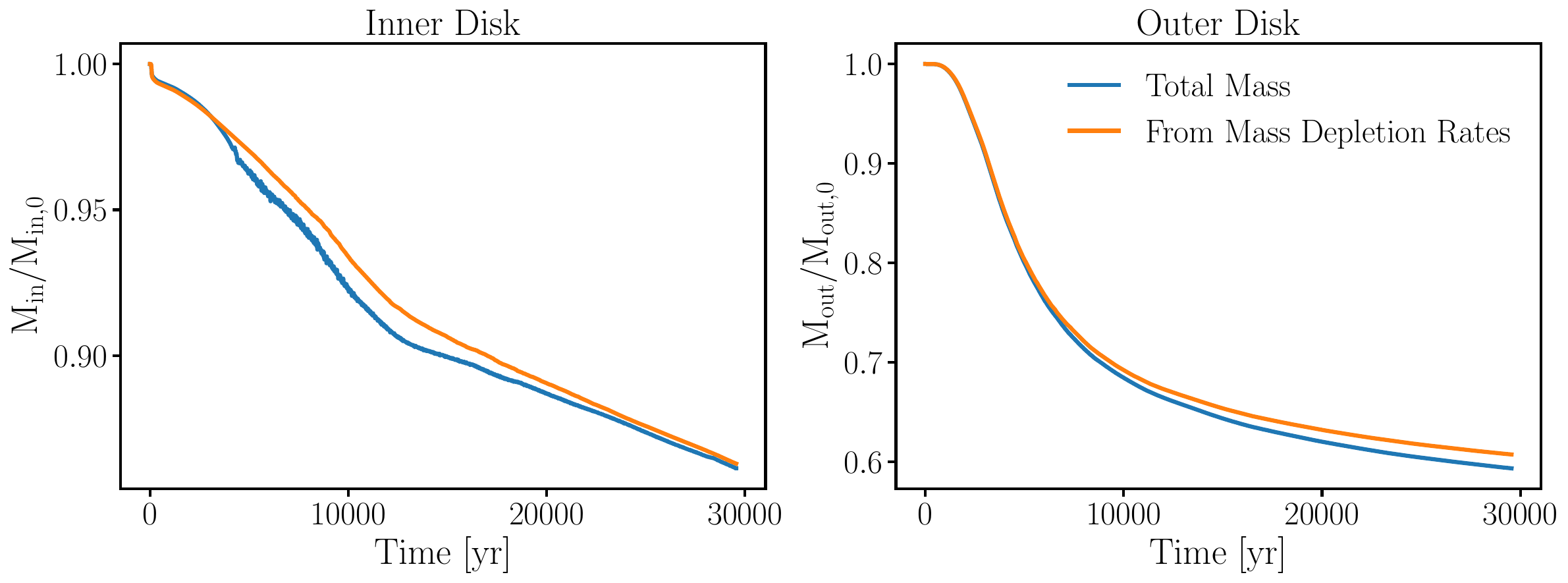}
    \caption{\textbf{Evolution of the disk mass in the inner (left) and outer disks (right) as a function of time in our nominal hydrodynamical simulation, including Jupiter.} The mass is normalized by the initial mass in the respective region. The inner region is defined as the region where  $r<5.4$~au and the outer disk as $r>5.4$~au. The blue curve is the same as the blue curve in Figure~\ref{fig:inner_disk_mass}, while the orange curve represents the mass in each region when integrating the accretion rates at the inner and outer boundaries of the disk. The main purpose of this figure is to show that the inner and outer disks are efficiently disconnected by Jupiter.}
    \label{fig:inner_outer_gas_mass}
\end{figure}

Due to the adopted open boundary conditions in our hydrodynamical simulations, the total gas mass in both the inner and outer disk decreases over time as gas is lost through the inner and outer edges. Supplementary figure~\ref{fig:inner_outer_gas_mass} illustrates the evolution of the disk mass in our nominal simulation. The blue curve represents the mass obtained from the \textsc{FARGO3D} simulations, while the orange curve shows the mass calculated from the instantaneous mass accretion rates at the respective disk edges using the following expression:
\begin{align} M_X(t) &= M_{X,0} - \int_{0}^t\dot{M}X(t)dt = M_0 - \int_{0}^t\int_X\Sigma_\text{gas}(r, t)v_r(r, t)drdt, \end{align}
where $X$ denotes either the inner or outer disk region. The negligible difference (within a few percent) between the two curves indicates that the mass loss from these regions is primarily due to outflows at the disk boundaries, with minimal mass exchange between them. This shows that Jupiter acts as an effective barrier between the inner and outer solar system.

\clearpage 

\end{document}